\RequirePackage[l2tabu, orthodox]{nag}
\RequirePackage{snapshot}

\documentclass[9pt,onecolumn]{extarticle}
\makeatletter
\if@twocolumn
  \usepackage[dvips,letterpaper,top=0.5in, bottom=0.5in, left=0.75in, right=0.5in,includefoot,heightrounded]{geometry}
\else
  \usepackage[dvips,letterpaper,margin=1in,includefoot,heightrounded]{geometry}
\fi

\usepackage{srcltx}

\usepackage{amsmath}
\usepackage{amssymb,amsfonts}

\usepackage{abstract}

\usepackage{graphicx}
\usepackage[usenames,dvipsnames,svgnames,x11names]{xcolor}
\usepackage{subfigure}

\usepackage{booktabs}

\usepackage{setspace}
\usepackage{flushend}

\usepackage{multicol}

\usepackage{cite}

\usepackage{enumerate}

\usepackage{multirow}

\usepackage[short,12hr]{datetime}

\usepackage{hyperref}
\usepackage[hyperpageref]{backref}

\makeatletter

\newcommand{\printtitle}{%
\makeatletter
\if@twocolumn

\twocolumn[%
  \maketitle
  \begin{onecolabstract}
    \myabstract
  \end{onecolabstract}
  \begin{center}
    \small
    \textbf{Keywords}
    \\\medskip
    \mykeywords
  \end{center}
  \bigskip
]
\saythanks
\else
  \maketitle
  \begin{onecolabstract}
    \myabstract
  \end{onecolabstract}
  \begin{center}
    \small
    \textbf{Keywords}
    \\\medskip
    \mykeywords
  \end{center}
  \bigskip
  \onehalfspacing
\fi
\makeatother
}

\title{A Multiparametric Class of Low-complexity Transforms for Image and Video Coding}

\author{
Diego Ramos Canterle%
\thanks{D. R. Canterle is with
the Instituto de Matem\'atica e Estat\'{\i}stica, Universidade de S\~ao Paulo, S\~ao Paulo, Brazil;
and
was with the
Programa de P\'os-Gradua\c{c}\~ao em Engenharia El\'etrica, Universidade Federal de Pernambuco, Recife, Brazil, e-mail: \url{diegocanterle@gmail.com}
}
\and
Thiago L. T. da Silveira%
\thanks{T. L. T. da Silveira
is with the Centro de Ci\^encias Computacionais, Universidade Federal do Rio Grande, Rio Grande, Brazil, e-mail: \url{tltsilveira@furg.br}}
\and
F\'abio Mariano Bayer%
\thanks{
F. M. Bayer is with the
Departamento de Estat\'{\i}stica and LACESM, Universidade Federal de Santa Maria, Santa Maria, RS, Brazil, e-mail: \url{bayer@ufsm.br}}
\and
Renato~J.~Cintra%
\thanks{R. J. Cintra is with the
Signal Processing Group, Departamento de Estat\'{\i}stica, Universidade Federal de Pernambuco, Recife, Brazil;
and
the
Department of Electrical and Computer Engineering, University of Calgary, Calgary, AB, Canada., e-mail: \url{rjdsc@de.ufpe.br}}
}

\date{\today\ @ \currenttime}

\date{}

\newcommand{\myabstract}{%
Discrete transforms play an important role in many signal processing applications, and low-complexity alternatives for classical transforms became popular in recent years. Particularly, the discrete cosine transform (DCT) has proven to be convenient for data compression, being employed in well-known image and video coding standards such as JPEG,
			H.264, and the recent high efficiency video coding (HEVC).
			In this paper, we introduce a new class of low-complexity 8-point DCT approximations based on a series of works published by Bouguezel, Ahmed and Swamy. Also, a  multiparametric fast algorithm that encompasses both known and novel transforms is derived.
			We select the best-performing DCT approximations after solving a multicriteria optimization problem, and submit them to a scaling method for obtaining larger size transforms.
			We assess these DCT approximations in both JPEG-like image compression and video coding experiments.
			We show that the optimal DCT approximations present compelling results in terms of coding efficiency and image quality metrics, and require only few addition or bit-shifting operations, being suitable for low-complexity and low-power systems.
}

\newcommand{\mykeywords}{%
Approximate transforms,
Arithmetic complexity,
Discrete cosine transform,
Image compression,
Video coding.
}

\begin{document}

\printtitle

\section{Introduction}

	The discrete cosine transform (DCT) \cite{Ahmed1974} is a fundamental tool in the digital signal processing field \cite{Ahmed1975,Oppenheim2009}.
	More precisely, the DCT is an asymptotic data-independent approximation
	for the optimal Karhunen-Loève transform (KLT)
	\cite{Efros2004}
	when the input signal
	can be modeled as a  first-order Markovian process, and the signal correlation coefficient tends to the unit ($ \rho \rightarrow 1 $) \cite{Cla1981}.
	Natural images are signals that belong to this statistical class \cite{Gonzalez2008}.

	The
	DCT has been
	successfully employed in well-known image and video coding standards, like
	JPEG \cite{Home}, H.264 \cite{Wiegand2003}
	and the recent high efficiency video coding (HEVC)
	\cite{hevc1}.
	Several DCT methods, such as \cite{Chen1977,Lee1984,Loef1989,Feig1992},
led to block-based transformations equipped with fast algorithms
that are capable of acceptable computational burden and
are widely adopted for efficient encoding of both images and video sequences.

    Nevertheless,  due to the irrational quantities in the DCT formulation, exact transformations might entail realizations
that require relatively demanding arithmetic schemes \cite{Brita2007},
such as floating-point or large integer fixed-point systems.
    Such constraint can preclude the applicability of the exact DCT computation in extremely low-power and real-time systems \cite{Wang2011, Saponara2012, bas2013}, such as~\cite{Harize2013,Kouadria2017}.
	In this context, approximate transforms
	can be efficiently implemented using only addition and bit-shifting operations
	\cite{bas2013,Haweel2001,Bayer2012,cb2011,Tablada2015,bas2011}.
	In fact, there are several methods in literature that focus on finding a good compromise between coding efficiency and computational cost \cite{bas2013,Bayer2012,cb2011,Tablada2015,bas2011,multibeam2012,Potluri2013,Bayer2010,bayer201216pt,Bayer2013,cintra2014low,bas2008,bas2008b,bas2009,	bas2010,Oliveira2019,Coutinho2017,Ezhilarasi2018,Almurib2018,Zhang2019,Huang2019,Zidani2019,Brahimi2020}.
    Classically, special attention was given to 8-point approximate transforms, since this particular blocklength is employed in both JPEG and H.264 coding standards.
    Nowadays, transforms of larger sizes, such as $N=16, 32$ are also required to cope with high-resolution video coding \cite{hevc1,Oliveira2019,Ohm2012}.

	A number of works proposing 8-point multiplication-free transforms in the context of image compression was introduced by Bouguezel-Ahmad-Swamy (BAS) \cite{bas2013,bas2011,bas2008,bas2008b,bas2009,bas2010}.
In particular, we emphasize the method described in~\cite{bas2011}
which employs a single parameter approximation for the 8-point DCT.
We aim at significantly extending such parameter-based approach.

	In this paper, we propose a new multiparametric class of low-complexity 8-point DCT approximations that
	encompasses the BAS transforms,
	and present the underlying fast algorithm.
	The obtained DCT approximations in the proposed class of low-complexity transforms are sought to be assessed and screened through an optimization process considering proximity measures relative to the exact DCT and coding efficiency metrics.
	Then, the best-performing transforms are submitted to a scaling method for obtaining 16- and 32-point DCT approximations,
aiming at the application in
recent image and video encoders~\cite{hevc1,Sullivan2012}.

	The rest of the paper is organized as follows.
	Section~\ref{sec:approximation} presents
	the mathematical formulation of the new class of DCT approximations.
	In Section~\ref{section-multicriteria},
	we explain the proposed
	multicriteria optimization scheme, and show the resulting 8-point low-complexity transforms.
	Section~\ref{section-scaled} introduces novel 16- and 32-point DCT approximations generated by the scaling method from \cite{Jrid2015}.
	Sections~\ref{ss:imagecoding} and~\ref{section-video} present, respectively, image and video coding experiments.
	Finally, Section~\ref{section-conclusion} concludes this work.

	\section{Multiparametric DCT approximations} \label{sec:approximation}

    In this section,
we review the DCT
and introduce the mathematical formulation for the proposed class of low-complexity transforms.
Computational complexity
and orthogonality property
are derived and discussed.

	\subsection{A review about the DCT} \label{ssec:preliminaries}

	Let $ \mathbf{C}_N $ be the transformation matrix related to the $N$-point DCT, for which the elements are given by \cite{Ahmed1974}
	\begin{align*}
			c_{i,j}=\alpha_{i}\cos\left(\dfrac{\pi i(2j+1)}{2N} \right),
	\end{align*}
	where
	$i,j=0,1,\ldots,N-1$,
	and
\[
\alpha_k =
\begin{cases}
1/\sqrt{N}, & \text{if $k=0$,} \\
\sqrt{2/N}, & \text{if $k>0$.}
\end{cases}
\]

Considering the standard approach for splitting images into disjoint sub-blocks \cite{Wallace1992}, the blockwise forward and inverse
two-dimensional
	DCT
	transformation are given,
	respectively, by
	\cite{Ahmed1974}
	\vspace{-0.1cm}
	\begin{align}
	\label{eq:dctfwd}
	\mathbf{B}=\mathbf{C}_N \cdot \mathbf{A} \cdot \mathbf{C}_N^{\top}
	\end{align}
	and
	\begin{align}
	\label{eq:dctbwd}
    \mathbf{A}=\mathbf{C}_N^{\top} \cdot \mathbf{B} \cdot \mathbf{C}_N,
	\end{align}
	where $ \mathbf{A} $ and $ \mathbf{B} $
	represent the input and transformed $N\times N$ signals, respectively.
	Note that because $ \mathbf{C}_N $ is orthogonal, the inverse transformation is immediately obtained by matrix transposition.
In practical terms, orthogonality implies that
the both forward and inverse  transformations
share similar realizations
\cite{Ahmed1975}.

Hereafter, we use the notation $\mathbf{\hat{C}}$ for referring to a given \emph{DCT approximation}. Generally, DCT approximations can be written as $\mathbf{\hat{C}} = \mathbf{S} \cdot \mathbf{{T}}$, where $\mathbf{S}$ is a diagonal scaling matrix and $\mathbf{{T}}$ is a low-complexity matrix whose entries are in the set $ \mathcal{C} = \{ 0, \pm 0.5, \pm 1, \pm 2 \} $.
Details are given in Section~\ref{ssec:ortho}.

	\subsection{Parametrization}

Most BAS low-complexity transforms
can be understood as variations of the signed DCT (SDCT) \cite{Haweel2001}
according to judicious changes in the matrix entries.
The SDCT can be obtained for any transform size $N$ by applying the signum function to all the entries of $\mathbf{C}_N$.
The resulting transformation matrix contains only elements in the set $\{\pm1\}$ and can be implemented using only additions, i.e., multiplications or bit-shifting operations are not required.
However, as a drawback, the SDCT lacks orthogonality for $N\neq 4$.

Based on the 8-point BAS orthogonal transforms
reported in the literature \cite{bas2013,bas2011,bas2008,bas2009,bas2010},
we aim at proposing a parametrization capable of encompassing such DCT approximations.
To attain the sought
multiparametric class of BAS-based transforms,
we perform entry-wise comparisons
of the considered BAS transforms
keeping common matrix blocks and
parametrizing the variations.
Thus the proposed multiparametric class
of DCT approximations based on BAS transforms is given by
the following expression:
	\begin{align}\label{E:P8}
	\mathbf{T}(\mathbf{a})=
	\begin{bmatrix}
	1 & 1 & 1 & 1 & 1 & 1 & 1 & 1 \\
	1 & 1 & a_1 & a_1 & -a_1 & -a_1 & -1 & -1 \\
	1 & a_2 & -a_2 & -1 & -1 & -a_2 & a_2 & 1 \\
	a_1 & a_3 & -a_4 & -a_1 & a_1 & a_4 & -a_3 & -a_1 \\
	1 & -1 & -1 & 1 & 1 & -1 & -1 & 1 \\
	a_5 & -a_5 & -a_1 & a_6 & -a_6 & a_1 & a_5 & -a_5 \\
	a_2 & -1 & 1 & -a_2 & -a_2 & 1 & -1 & a_2 \\
	a_7 & -a_6 & a_1 & -a_8 & a_8 & -a_1 & a_6 & -a_7
	\end{bmatrix},
	\end{align}
where
$ \mathbf{a} = [a_1 \, a_2 \, a_3 \, a_4 \, a_5 \, a_6 \, a_7 \, a_8]^\top \in \mathbb{R}^8$  is
	the parameter vector for matrix generation.
	Depending on the values of $ \mathbf{a} $ we can find different transforms as special cases,
	such as those proposed in \cite{bas2013,bas2011,bas2008,bas2009,bas2010}.
	In order to guarantee low-complexity transforms we shall consider that $a_i \in \mathcal{C}$, for $i=1,\ldots,8$.

	Note that reducing the computational complexity is only one possible requirement. Since we are interested in exploring BAS-based transforms in the context of image and video coding, it is also important to attain good compaction properties \cite{Brita2007}. BAS transforms do have high coding gain and transform efficiency measurements \cite{bas2008, bas2009, bas2010, Brahimi2020}, and we expect to find novel transforms sharing the same properties (cf. Section \ref{section-multicriteria}).

	\subsection{Fast algorithm}

    Following similar approach as in \cite{Bayer2012,cb2011,Tablada2015,Oliveira2019}, we sparsely factorize the low-complexity matrix given in Equation~\eqref{E:P8} as
\begin{align*}
	\mathbf{T}(\mathbf{a}) = \mathbf{P} \cdot \mathbf{K}(\mathbf{a}) \cdot \mathbf{A_2} \cdot \mathbf{A_1},
\end{align*}
where
	\begin{align*}
	\mathbf{A_1}=
	\begin{bmatrix}
	1 & 0 & 0 & 0 & 0 & 0 & 0 & 1 \\
	0 & 1 & 0 & 0 & 0 & 0 & 1 & 0 \\
	0 & 0 & 1 & 0 & 0 & 1 & 0 & 0 \\
	0 & 0 & 0 & 1 & 1 & 0 & 0 & 0 \\
	0 & 0 & 0 & 1 & -1 & 0 & 0 & 0 \\
	0 & 0 & 1 & 0 & 0 & -1 & 0 & 0 \\
	0 & 1 & 0 & 0 & 0 & 0 & -1 & 0 \\
	1 & 0 & 0 & 0 & 0 & 0 & 0 & -1
	\end{bmatrix},
	\end{align*}
	\begin{align*}
	\mathbf{A_2}=
	\begin{bmatrix}
	1 & 0 & 0 & 1 & 0 & 0 & 0 & 0 \\
	0 & 1 & 1 & 0 & 0 & 0 & 0 & 0 \\
	0 & 1 & -1 & 0 & 0 & 0 & 0 & 0 \\
	1 & 0 & 0 & -1 & 0 & 0 & 0 & 0 \\
	0 & 0 & 0 & 0 & 1 & 0 & 0 & 0 \\
	0 & 0 & 0 & 0 & 0 & 1 & 0 & 0 \\
	0 & 0 & 0 & 0 & 0 & 0 & 1 & 0 \\
	0 & 0 & 0 & 0 & 0 & 0 & 0 & 1
	\end{bmatrix},
	\end{align*}
	\begin{align*}
	\mathbf{K}(\mathbf{a})=
	\begin{bmatrix}
	1 & 1 & 0 & 0 & 0 & 0 & 0 & 0 \\
	1 & -1 & 0 & 0 & 0 & 0 & 0 & 0 \\
	0 & 0 & a_2 & 1 & 0 & 0 & 0 & 0 \\
	0 & 0 & -1 & a_2 & 0 & 0 & 0 & 0 \\
	0 & 0 & 0 & 0 & a_1 & a_1 & 1 & 1 \\
	0 & 0 & 0 & 0 & a_6 & -a_1 & -a_5 & a_5 \\
	0 & 0 & 0 & 0 & -a_1 & -a_4 & a_3 & a_1 \\
	0 & 0 & 0 & 0 & -a_8 & a_1 & -a_6 & a_7
	\end{bmatrix},
	\end{align*}
	and
	\begin{align*}
	\mathbf{P}=
	\begin{bmatrix}
	1 & 0 & 0 & 0 & 0 & 0 & 0 & 0 \\
	0 & 0 & 0 & 0 & 1 & 0 & 0 & 0 \\
	0 & 0 & 1 & 0 & 0 & 0 & 0 & 0 \\
	0 & 0 & 0 & 0 & 0 & 0 & 1 & 0 \\
	0 & 1 & 0 & 0 & 0 & 0 & 0 & 0 \\
	0 & 0 & 0 & 0 & 0 & 1 & 0 & 0 \\
	0 & 0 & 0 & 1 & 0 & 0 & 0 & 0 \\
	0 & 0 & 0 & 0 & 0 & 0 & 0 & 1
	\end{bmatrix}.
	\end{align*}

\begin{figure}
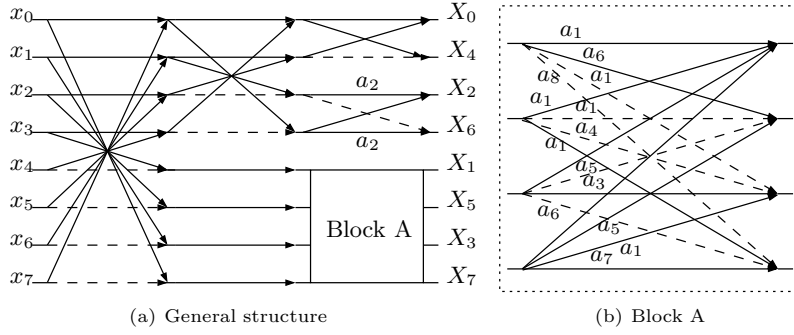

	\centering
	\subfigure[General structure]{\input{figures/fig1diagram.pstex_t}}\qquad
	\subfigure[Block A]{\input{figures/fig1block.pstex_t}}
	\caption{
	SFG for $\mathbf{T(a)}$.
  Input data $x_i, i = 0, 1, \ldots, 7,$ relates to output $X_j, j = 0, 1, \ldots, 7,$ according to $X = \mathbf{T(a)}\cdot x$. Dashed arrows represent multiplications by $-1$.}
	\label{F:f1}
	\end{figure}

Note that only matrix $\mathbf{K(a)}$ depends on the parameter vector $\mathbf{a}$.
The signal flow graph (SFG) for the proposed fast algorithm is shown in Figure~\ref{F:f1}.
	We discuss in detail the computational complexity associated to $\mathbf{T(a)}$ in the following.

	\subsection{Computational complexity}\label{SS:cc}

Matrices $\mathbf{A_1}$ and $\mathbf{A_2}$ contribute with additions only.
Matrix $\mathbf{P}$ represents a permutation,
which is multiplierless and corresponds to wiring in terms of circuit implementation.
	The computational complexity of matrix $\mathbf{K(a)}$ depends on the parameter vector $\mathbf{a}$.
	Here, we consider that the elements of $\mathbf{a} $ are in the set $\mathcal{C}$,
	so that only additions and bit-shifting operations are required for implementing the fast algorithms of $\mathbf{T(a)}$
(see Figure~\ref{F:f1}).

	Thus, the
	number of additions $\mathcal{A} (\mathbf{a})$ and bit-shifting operations $\mathcal{S} (\mathbf{a})$ required for implementing $\mathbf{T(a)}$ are given, respectively,
	by
	\small
	\begin{align*}
\mathcal{A}(\mathbf{a}) = 28 - \sum^8_{i=1}w_i\mathcal{I}_{\{ 0 \}}(a_i)   \quad\text{and}\quad
\mathcal{S}(\mathbf{a}) = \sum^8_{i=1}w_i\mathcal{I}_{\left\lbrace  \frac{1}{2},2 \right\rbrace}(a_i),
	\end{align*}
	\normalsize
where $\mathbf{w}=[6\, 2\, 1\, 1\, 2\, 2\, 1\, 1]^{\top}$,
and
$\mathcal{I}_{X}(x) = 1$,
if $x \in X$,
otherwise it returns zero.

These equations compute additive and bit-shift complexity considering the fast algorithm
presented in Figure \ref{F:f1}.
This fast algorithm is general and the arithmetic complexity can be further reduced depending on the parameter vector $\mathbf{a}$.
Specific combinations of the scalars within $\mathbf{a}$ lead to simplified versions of $ \mathbf{A_2} $ and $ \mathbf{K}( \mathbf{a}) $ that require fewer computations. We list all nine restrictions and their modified addition and bit-shifting counts below.
	\begin{enumerate}
	\small
\item If $|a_1| = |a_4| = |a_6| = |a_8|$, then
\begin{align*}
	\mathcal{A}(\mathbf{a}) = 26 - \sum^8_{i=1}w_i\mathcal{I}_{\{ 0 \}}(a_i)\quad\text{and}\quad
	\mathcal{S}(\mathbf{a}) = \sum^8_{i=1}w_i\mathcal{I}_{\left\lbrace  \frac{1}{2},2 \right\rbrace}(a_i),
\end{align*}
where $\mathbf{w}=[6\, 2\, 1\, 0\, 2\, 0\, 1\, 0]^{\top}$;

\item If $(|a_1| = |a_3| = 1) \wedge (|a_5| = |a_6| = |a_8|)$, then
\begin{align*}
	\mathcal{A}(\mathbf{a}) = 26 - \sum^8_{i=1}w_i\mathcal{I}_{\{ 0 \}}(a_i)\quad\text{and}\quad
	\mathcal{S}(\mathbf{a}) = \sum^8_{i=1}w_i\mathcal{I}_{\left\lbrace  \frac{1}{2},2 \right\rbrace}(a_i),
\end{align*}
where $\mathbf{w}=[0\, 2\, 0\, 1\, 3\, 0\, 1\, 0]^{\top}$;

\item If $(|a_1| = 1) \wedge (|a_5| = |a_6|) \wedge (|a_7| = |a_8|)$, then
\begin{align*}
	\mathcal{A}(\mathbf{a}) = 26 - \sum^8_{i=1}w_i\mathcal{I}_{\{ 0 \}}(a_i)\quad\text{and}\quad
	\mathcal{S}(\mathbf{a}) = \sum^8_{i=1}w_i\mathcal{I}_{\left\lbrace  \frac{1}{2},2 \right\rbrace}(a_i),
\end{align*}
where $\mathbf{w}=[0\, 2\, 1\, 1\, 3\, 0\, 1\, 0]^{\top}$;

\item If $(|a_1| = |a_5| = |a_6| = 1) \wedge (|a_3| = |a_4|)$, then
\begin{align*}
	\mathcal{A}(\mathbf{a}) = 26 - \sum^8_{i=1}w_i\mathcal{I}_{\{ 0 \}}(a_i)\quad\text{and}\quad
	\mathcal{S}(\mathbf{a}) = \sum^8_{i=1}w_i\mathcal{I}_{\left\lbrace  \frac{1}{2},2 \right\rbrace}(a_i),
\end{align*}
where $\mathbf{w}=[0\, 2\, 1\, 0\, 0\, 0\, 1\, 1]^{\top}$;

\item If $|a_1| = |a_4| = |a_5| = |a_7| = 1$, then
\begin{align*}
	\mathcal{A}(\mathbf{a}) = 26 - \sum^8_{i=1}w_i\mathcal{I}_{\{ 0 \}}(a_i)\quad\text{and}\quad
	\mathcal{S}(\mathbf{a}) = \sum^8_{i=1}w_i\mathcal{I}_{\left\lbrace  \frac{1}{2},2 \right\rbrace}(a_i),
\end{align*}
where $\mathbf{w}=[0\, 2\, 1\, 0\, 0\, 2\, 0\, 1]^{\top}$;

\item If $(|a_1| = |a_3|) \wedge (|a_6| = |a_7|)$, then
\begin{align*}
	\mathcal{A}(\mathbf{a}) = 26 - \sum^8_{i=1}w_i\mathcal{I}_{\{ 0 \}}(a_i)\quad\text{and}\quad
	\mathcal{S}(\mathbf{a}) = \sum^8_{i=1}w_i\mathcal{I}_{\left\lbrace  \frac{1}{2},2 \right\rbrace}(a_i),
\end{align*}
where $\mathbf{w}=[6\, 2\, 0\, 1\, 1\, 2\, 0\, 1]^{\top}$;

\item If $|a_1| = |a_3| = |a_4| = |a_6| = |a_7| = |a_8|$, then
\begin{align*}
	\mathcal{A}(\mathbf{a}) = 24 - \sum^8_{i=1}w_i\mathcal{I}_{\{ 0 \}}(a_i)\quad\text{and}\quad
	\mathcal{S}(\mathbf{a}) = \sum^8_{i=1}w_i\mathcal{I}_{\left\lbrace  \frac{1}{2},2 \right\rbrace}(a_i),
\end{align*}
where $\mathbf{w}=[6\, 2\, 0\, 0\, 1\, 0\, 0\, 0]^{\top}$;

\item If $|a_1| = |a_3| = |a_4| = |a_5| = |a_6| = |a_7| = |a_8| = 1$, then
\begin{align*}
	\mathcal{A}(\mathbf{a}) = 24 - \sum^8_{i=1}w_i\mathcal{I}_{\{ 0 \}}(a_i)\quad\text{and}\quad
	\mathcal{S}(\mathbf{a}) = \sum^8_{i=1}w_i\mathcal{I}_{\left\lbrace  \frac{1}{2},2 \right\rbrace}(a_i),
\end{align*}
where $\mathbf{w}=[0\, 2\, 0\, 0\, 0\, 0\, 0\, 0]^{\top}$;

\item If $(|a_1| = |a_5| = |a_6| = 1) \wedge (|a_3| = |a_4|) \wedge (|a_7| = |a_8|)$, then
\begin{align*}
	\mathcal{A}(\mathbf{a}) = 24 - \sum^8_{i=1}w_i\mathcal{I}_{\{ 0 \}}(a_i)\quad\text{and}\quad
	\mathcal{S}(\mathbf{a}) = \sum^8_{i=1}w_i\mathcal{I}_{\left\lbrace  \frac{1}{2},2 \right\rbrace}(a_i),
\end{align*}
where $\mathbf{w}=[0\, 2\, 1\, 0\, 0\, 0\, 1\, 0]^{\top}$.
\end{enumerate}

Above expressions show that
$ 16 \leq \mathcal{A}(\mathbf {a}) \leq 28 $
and
$ 0 \leq \mathcal{S} (\mathbf{a}) \leq 16 $.

	\subsection{Orthogonality and orthonormality} \label{ssec:ortho}

Discrete transforms are often required to be orthogonal~\cite{Brita2007,Leng2004}.
One of the reasons is the fact that orthogonality
ensures that the good mathematical properties of
the forward transformation
are
transferred to the inverse operation.

Here,
a matrix
	 $ \textbf{T} $ is said to be orthogonal
	if
	$ \textbf{T} \cdot \textbf{T}^{\top}$ is a diagonal matrix.
If $ \textbf{T} \cdot \textbf{T}^{\top}$
	is the identity matrix
	then $ \textbf{T} $ is
referred to as orthonormal.
Considering the proposed parametrization (Equation \eqref{E:P8}),
we have:
	\begin{align*}
	\mathbf{T(\mathbf{a})} \cdot \mathbf{T(\mathbf{a})}^\top =
	\begin{bmatrix}
	8 & 0 & 0 & 0 & 0 & 0 & 0 & 0 \\
	0 & \tau_1 & 0 & \tau_6 & 0 & \tau_7 & 0 & \tau_8 \\
	0 & 0 & \tau_2 & 0 & 0 & 0 & 0 & 0 \\
	0 & \tau_6 & 0 & \tau_3 & 0 & \tau_9 & 0 & \tau_{10} \\
	0 & 0 & 0 & 0 & 8 & 0 & 0 & 0 \\
	0 & \tau_7 & 0 & \tau_9 & 0 & \tau_4 & 0 & \tau_{11} \\
	0 & 0 & 0 & 0 & 0 & 0 & \tau_2 & 0 \\
	0 & \tau_8 & 0 & \tau_{10} & 0 & \tau_{11} & 0 & \tau_5
	\end{bmatrix},
	\end{align*}
	where $\tau_1 = 4a_1^2+4$, $\tau_2 = 4a_2^2+4$, $\tau_3 = 4a_1^2+2a_3^2+2a_4^2$,
	$\tau_4 = 2a_6^2+4a_5^2+2a_1^2$, $\tau_5 = 2a_8^2+2a_7^2+2a_6^2+2a_1^2$,
	$\tau_6 = 2a_1-2a_1^2+2a_3-2a_1a_4$, $\tau_7 = 2a_1a_6-2a_1^2$,
	$\tau_8 = 2a_1^2-2a_6+2a_7-2a_1a_8$, $\tau_9 = 2a_1a_4+2a_1a_5-2a_3a_5-2a_1a_6$,
	$\tau_{10} = 2a_1a_8+2a_1a_7-2a_3a_6-2a_1a_4$ and
    $\tau_{11} = 2a_5a_7+2a_5a_6-2a_1^2-2a_6a_8$.
Matrix  $\mathbf{T(a)}$
is orthogonal,
if its entries
	$\tau_6$, $\tau_7$, $\tau_8$, $\tau_9$, $\tau_{10}$,
	and $\tau_{11}$
are equal to zero.
Therefore,
the following conditions must hold true
to ensure orthogonality:
	\begin{align*}
	\left\{ \begin{array}{ll}
	2a_1-2a_1^2+2a_3-2a_1 a_4 &= 0, \\
	2a_1a_6-2a_1^2 &= 0, \\
	2a_1^2-2a_6+2a_7-2a_1 a_8 &= 0, \\
	2a_1 a_4+2a_1 a_5-2a_3 a_5-2a_1 a_6 &= 0, \\
	2 a_1 a_8+2a_1 a_7-2a_3 a_6-2a_1 a_4 &= 0, \\
	2a_5 a_7+2a_5 a_6-2a_1^2-2a_6 a_8 &= 0.
	\end{array} \right.
	\end{align*}

Orthonormality can be obtained by means of polar decomposition \cite{Bayer2012,cb2011,multibeam2012}.
An
orthonormal DCT approximation
 $\hat{\mathbf{C}}(\mathbf{a})$
is given by~\cite{Cintra2011}
	\begin{align*}
	\hat{\mathbf{C}}(\mathbf{a}) = \mathbf{S(\mathbf{a})}\cdot \mathbf{T(\mathbf{a})},
	\end{align*}
	where
	\begin{align*}
	\mathbf{S(\mathbf{a})} = \sqrt{\left[\mathbf{T(\mathbf{a})} \cdot \mathbf{T(\mathbf{a})}^\top\right]^{-1}},
	\end{align*}
	and $ \sqrt{\cdot} $ denotes the matrix square root \cite{Higham1987}.
Thus,
we
have that
	\small
	\begin{align*}
	\mathbf{S(\mathbf{a})} = \operatorname{diag}\left( \dfrac{1}{2\sqrt{2}}, \dfrac{1}{\sqrt{\tau_1}}, \dfrac{1}{\sqrt{\tau_2}}, \dfrac{1}{\sqrt{\tau_3}}, \dfrac{1}{2\sqrt{2}}, \dfrac{1}{\sqrt{\tau_4}}, \dfrac{1}{\sqrt{\tau_2}}, \dfrac{1}{\sqrt{\tau_5}} \right).
	\end{align*}
	\normalsize
However,
in the context of image and video coding, not only $\mathbf{T(\mathbf{a})}$ is said to be of low-complexity but also $\mathbf{\hat{C}(\mathbf{a})}$ because the scaling matrix $\mathbf{S(\mathbf{a})} $ can be merged in the quantization step \cite{Bayer2012,cb2011,bas2011,bayer201216pt,bas2008,bas2009,Leng2004}.
Therefore,
$\mathbf{S(\mathbf{a})} $
does not contribute with any computational complexity.

\section{Multicriteria optimization}

\label{section-multicriteria}

In this section,
we employ multicriteria optimization
for finding
DCT approximations
according to the mathematical formalism
discussed in the previous section.
Frequently,
two types of metrics are used
for assessing a given approximate DCT \cite{Tablada2015}:
proximity measures
and
coding measures.
Proximity measures
assess
how close this transform is to the exact DCT
in a Euclidean sense,
implying the low measurements are sought.
On the other hand,
coding measures
aim at capturing how good
a transformation is in terms of energy compaction properties;
high values of coding are desirable.
Total error energy~\cite{cb2011} and mean square error (MSE)~\cite{Brita2007}
are adopted as proximity measures;
whereas
coding gain~\cite{jayant1984digital} and transform efficiency~\cite{Brita2007}
are selected for coding measurements.
Moreover,
the number of additions and number of bit-shift operations
are also considered as figures of metric
for complexity
and
are sought to be minimized.
The above discussion
entails the following
multicriteria minimization problem:
\begin{align}
\label{E:optim}
\min_\mathbf{a}
(
\epsilon(\mathbf{\hat C(a)}),
\operatorname{MSE}(\mathbf{\hat C(a)}),
-C_g^*(\mathbf{\hat C(a)}),
-\eta(\mathbf{\hat C(a)}),
\mathcal{A} (\mathbf {a}),
\mathcal{S} (\mathbf{a})
)
,
\end{align}
	where $ \epsilon(\cdot) $,
	$ \operatorname{MSE} (\cdot) $,
	$ C_g^* (\cdot) $,
	and $ \eta (\cdot) $
	compute, respectively,  the total error energy, the mean square error, the unified coding gain and the transform efficiency.
	When necessary, $\mathcal{A}(\mathbf {a})$ and $\mathcal{S} (\mathbf{a})$
	are computed according to the restrictions on the elements of $\mathbf{a}$ described in Section~\ref{SS:cc}.

For $a_i \in \mathcal{C}$,
$i=1,2,\ldots,8$,
	then there are $7^8 = 5{,}764{,}801 $
candidate matrices
in the search space of~\eqref{E:optim}.
Thus,
the above problem can be solved by means of exhaustive search
in contemporary computers.
The exhaustive search demanded four weeks of uninterrupted
calculations in a computer with an Intel Core i5 (3th generation)
processor equipped with 6GB of
RAM running \texttt{R} language~\cite{R2016} on Linux OS
(Ubuntu 16.04 LTS).

	Table~\ref{T:E8} presents
the obtained fifteen optimal solutions of
\eqref{E:optim}.
For the sake of simplicity,
we refer to the optimal 8-point DCT approximations
as
$\mathbf{\hat C}_{8,j}$, $j = 1, 2, \ldots, 15$.
Note that $\mathbf{\hat{C}}_{8,2}$, $\mathbf{\hat{C}}_{8,5}$,
and $\mathbf{\hat{C}}_{8,15}$
coincide
with literature results~\cite{bas2009,bas2008,bas2010}.
The transform $\mathbf{\hat{C}}_{8,1}$ was previously published in \cite{Brahimi2020,Oliveira2013b},
which consists of judiciously changing few entries of the matrix describes in~\cite{Brahimi2011}
leading to orthogonalization.
To the best of our knowledge,
the remaining
eleven matrices are novel results.

		\begin{table}
		\centering
		\caption{
		Optimal 8-point DCT approximations}
		\label{T:E8}
			\begin{tabular}{ccccccc}
				\hline
				$j$ &	$\mathbf{a}$ & Comment \\
				\hline
				1 & $[0\,	0\,	0\,	1\,	1\,	0\,	0\,	1]^{\top}$ & \cite{Brahimi2020,Oliveira2013b} \\
				2 & $[0\,	1\,	0\,	1\,	1\,	0\,	0\,	1]^{\top}$ & \cite{bas2009} \\
				3 & $[0\,	0\,	0\,	1\,	0.5\,	1\,	1\,	1]^{\top}$ & New transform\\
				4 & $[0\,	0\,	0\,	1\,	1\,	1\,	1\,	2]^{\top}$ & New transform \\
				5 & $[0\,	0.5\,	0\,	1\,	1\,	0\,	0\,	1]^{\top}$ & \cite{bas2008} \\
				6 & $[1\,	0\,	0\,	0\,	1\,	1\,	0\,	0]^{\top}$ & New transform \\
				7 & $[0\,	1\,	0\,	1\,	1\,	1\,	1\,	2]^{\top}$ & New transform \\
				8 & $[0\,	1\,	0\,	1\,	0.5\,	1\,	1\,	1]^{\top}$ & New transform \\
				9 & $[0\,	0.5\,	0\,	1\,	1\,	1\,	1\,	2]^{\top}$ & New transform \\
				10 & $[0\,	0.5\,	0\,	1\,	0.5\,	1\,	1\,	1]^{\top}$ & New transform \\
				11 & $[1\,	0\,	1\,	1\,	1\,	1\,	1\,	1\,]^{\top}$ & New transform \\
				12 & $[1\,	0.5\,	0\,	0\,	1\,	1\,	0\,	0]^{\top}$ & New transform \\
				13 & $[1\,	0\,	0.5\,	0.5\,	1\,	1\,	0.5\,	0.5]^{\top}$ & New transform \\
				14 & $[1\,	0.5\,	1\,	1\,	1\,	1\,	1\,	1]^{\top}$ & New transform \\
				15 & $[1\,	0.5\,	0.5\,	0.5\,	1\,	1\,	0.5\,	0.5]^{\top}$ & \cite{bas2010}\\
				\hline
			\end{tabular}
	\end{table}

Table~\ref{T:Avaliacao8}
displays
the
proximity and coding measurements,
as well as arithmetic complexity,
for the obtained approximations.
The best measurements are highlighted in boldface.
The approximation $ \mathbf{\hat{C}}_{8,15}$
attained the best results for proximity and coding metrics.
However, as a trade-off,
$ \mathbf{\hat{C}}_{8,15}$
has the highest computational complexity.
The approximation with the lowest computational cost is
$\mathbf{\hat{C}}_{8,1}$.
	Among the new
	transforms,
	$\mathbf{\hat{C}}_{8,9}$
	is worth-mentioning for its proximity to the exact DCT, and
	$\mathbf{\hat{C}}_{8,14}$
	and
	$\mathbf{\hat{C}}_{8,12}$,
	for the high coding gain and transform efficiency, respectively.

	\begin{table}
		\centering
		\caption{
		Results for the optimal 8-point DCT approximations.}
		\label{T:Avaliacao8}
			\begin{tabular}{ccccccc}
				\hline
				$j$ & $\epsilon(\cdot)$ &	MSE$(\cdot)$ &	$C_g^*(\cdot)$ &	$\eta(\cdot)$ & $\mathcal{A}(\cdot)$ & $\mathcal{S}(\cdot)$\\
				\hline
				1 & 6.85&	0.03&	7.91&	85.64&	\textbf{16}&	\textbf{0} \\
				2 & 6.85&	0.03&	7.91&	85.38&	18&	\textbf{0} \\
				3 & 5.79&	0.03&	7.91&	85.78&	18&	1 \\
				4 & 5.05&	0.03&	7.91&	85.51&	18&	1 \\
				5 & 5.93&	0.02&	8.12&   86.86&	18&	2 \\
				6 & 6.85&	0.03&	7.93&	85.80&	20&	\textbf{0} \\
				7 & 5.05&	0.03&	7.91&	85.25&	20&	1 \\
				8 & 5.79&	0.03&   7.91&	85.52&	20&	1 \\
				9 & 4.12&	\textbf{0.02}&	8.12&	86.73&	20&	3 \\
				10 & 4.87&	\textbf{0.02}&	8.12&	87.01&	20&	3 \\
				11 & 5.05&	\textbf{0.02}&	7.95&	85.58&	22&	\textbf{0} \\
				12 & 5.93&	\textbf{0.02}&	8.14&	87.02&	22&	2 \\
				13 & 5.02&	\textbf{0.02}&	8.12&	86.96&	22&	2 \\
				14 & 4.12&	\textbf{0.02}&	8.15&	86.79&	24&	2 \\
				15 & \textbf{4.09} & \textbf{0.02}&	\textbf{8.33}&	\textbf{88.22}&	24&	4 \\
				\hline
			\end{tabular}
	\end{table}

For further comparison,
we
show
the performance measurements
of
the following competing DCT
approximations that are out of our class of transformations defined in \eqref{E:P8}:
	FW$_5$ and FW$_6$ approximations from \cite{Tablada2015},
	the rounded discrete cosine transform (RDCT) \cite{cb2011},
	Lengwehasatit and Ortega level 1 approximation (LO) \cite{Leng2004},
	and
angle-based DCT approximation (ABM) \cite{Oliveira2019}.
We also list BAS transformations
\cite{bas2011,bas2009,bas2008,bas2013,bas2010}
that were not considered optimal according to our methodology
but they are special cases of our class of transformations.
The BAS transform proposed in \cite{bas2011} BAS$_5(a)$
is a uniparametric transform, where $a$ is the parameter.
Table~\ref{T:Comparacao8}
summarizes the measurements.

	\begin{table}
		\centering
	\setlength{\tabcolsep}{3pt}
		\caption{
		Results for competing 8-point DCT approximations.} \label{T:Comparacao8}
			\begin{tabular}{lcccccc}
				\hline
				Transform & $\epsilon(\cdot)$ &	MSE$(\cdot)$ &	$C_g^*(\cdot)$ &	$\eta(\cdot)$ & $\mathcal{A}(\cdot)$ & $\mathcal{S}(\cdot)$\\
				\hline
				BAS$_5(0)$ \cite{bas2011} & 26.86 & 0.07 & 7.91 & 85.64 & \textbf{16} & \textbf{0} \\
				FW$_6$ \cite{Tablada2015} & 3.32  & 0.02 & 6.05 & 83.08 & 18 & \textbf{0} \\
				BAS$_3$ \cite{bas2009} & 6.85  & 0.03 & 7.91 & 85.38 & 18 & \textbf{0} \\
				BAS$_5(1)$ \cite{bas2011} & 26.86 & 0.07 & 7.91 & 85.38 & 18 & \textbf{0} \\
				BAS$_5(1/2)$ \cite{bas2011} & 26.40 & 0.07 & 8.12 & 86.86 & 18 & 2 \\
				BAS$_1$ \cite{bas2008} & 5.93  & 0.02 & 8.12 & 86.86 & 18 & 2 \\
				FW$_5$ \cite{Tablada2015} & 7.41  & 0.05 & 7.58 & 83.08 & 20 & 10 \\
				RDCT \cite{cb2011}   & 1.79  & \textbf{0.01} & 8.18 & 87.43 & 22 & \textbf{0} \\
				BAS$_6$ \cite{bas2013} & 35.06 & 0.10 & 7.95 & 85.31 & 24 & \textbf{0} \\
				LO \cite{Leng2004} & \textbf{0.87} & \textbf{0.01} & 8.39 & 88.70 & 24 & 2 \\
				BAS$_4$ \cite{bas2010} & 4.09  & 0.02 & 8.33 & 88.22 & 24 & 4 \\
				ABM \cite{Oliveira2019} & 1.22  & \textbf{0.01} & \textbf{8.63} & \textbf{90.46} & 24 & 6 \\
				\hline
			\end{tabular}
	\end{table}

\begin{figure*}
		\centering
			\subfigure[Total energy error]{
				{\includegraphics[width=0.35\textwidth]{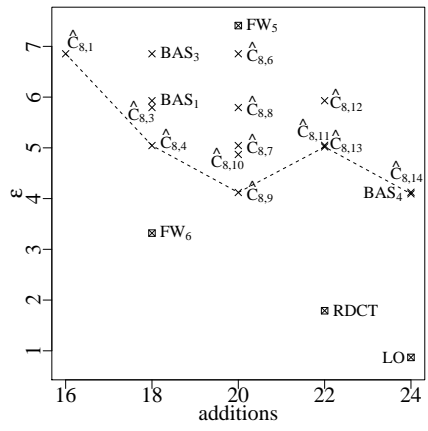}}\label{F:TEE8}}
				\qquad
			\subfigure[MSE]{
				{\includegraphics[width=0.35\textwidth]{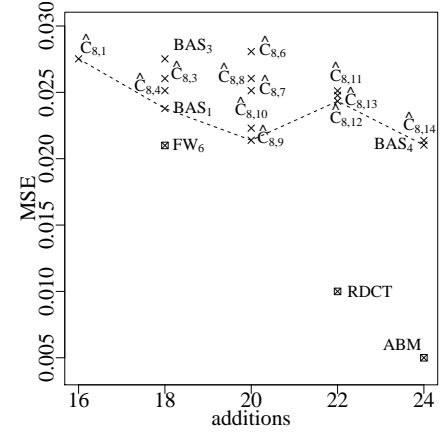}}\label{F:MSE8}}
				\qquad
			\subfigure[Unified coding gain]{
				{\includegraphics[width=0.35\textwidth]{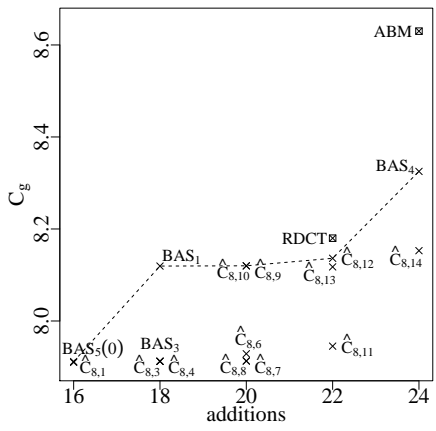}}\label{F:Cg8}}
				\qquad
			\subfigure[Transform efficiency]{
				{\includegraphics[width=0.35\textwidth]{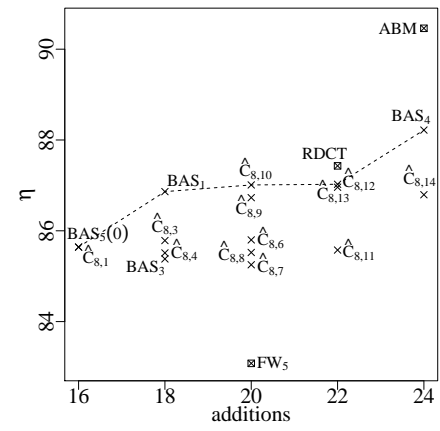}}\label{F:eta8}}
			\caption{Assessment plots for the proposed efficient approximations and competing methods. Dashed line represents the Pareto boundary for the transforms belonging to the proposed class.} \label{F:comparacao8}
	\end{figure*}

Figure~\ref{F:comparacao8} relates the computational complexity to the four metrics in Equation~\eqref{E:optim}
for all considered DCT approximations.
The dotted curves refer to the Pareto boundary which indicates the optimal transforms for each case~\cite{Miettinen1999}.
Considering transforms with 16~additions,
$\mathbf{\hat{C}}_{8,1}$ \cite{Oliveira2013b} achieves
the best measurements in terms of MSE and total error energy.
Approximation
$\mathbf{\hat{C}}_{8,1}$
presents identical results as the ones from the approximation BAS$_5(0)$
when coding gain and transform efficiency are considered.
Approximations
FW$_6$ and BAS$_1$ attain
the best results in terms of proximity to the exact DCT and coding efficiency, respectively, if a maximum of 18 additions is considered.
When considering transforms that require 20 additions,
	$\mathbf{\hat{C}}_{8,9}$
	performs better in terms of MSE, total energy error and coding gain.
	The highest transform efficiency is achieved by $\mathbf{\hat{C}}_{8,10}$.
	RDCT is the best-performing if compared to other transforms requiring 22 additions.
	Finally, for the case of 24 additions, LO transform achieves the minimum MSE, and ABM has the best results for the other metrics.
	Note that the transforms of our class of transformations requiring 16, 18, and 20 additions are usually the
	best-performing in terms of coding gain and transform efficiency.

\section{Scaled optimal transforms}

\label{section-scaled}

The widely popular HEVC standard adopts
larger size transforms to produce high-resolution video coding \cite{hevc1}.
In response,
16- and 32-point DCT approximations
have been proposed~\cite{bas2010,bas2013,Jrid2015,Silveira201660,Silveira201644,Oliveira2019}.
In this work,
we rely on the scaling method proposed by Jridi-Alfalou-Meher (JAM) \cite{Jrid2015}  which takes as input a low-complexity transform of size $N$ and generates
a transform of size $2N$.
We refer the reader to~\cite{Jrid2015} for more details about the scaling
procedure.

\subsection{Proposed 16-point DCT approximations}

Following the JAM method,
we obtain 16-point DCT approximations
based on the parametrization shown
in Equation~\eqref{E:P8},
resulting in:
\begin{align*}
\mathbf{T}_{16}(\mathbf{a}) =
\begin{bmatrix}
\mathbf{T}_{8}(\mathbf{a}) & \mathbf{0}_{8} \\
\mathbf{0}_{8} & \mathbf{T}_{8}(\mathbf{a})
\end{bmatrix}
\cdot
\begin{bmatrix}
\mathbf{I}_{8} &  \overline{\mathbf{I}}_{8} \\
\mathbf{I}_{8} & -\overline{\mathbf{I}}_{8}
\end{bmatrix}
\end{align*}
where
$\mathbf{0}_{8}$ is $8\times8$
a matrix of zeros;
$\mathbf{I}_{8}$
and
$\overline{\mathbf{I}}_{8}$
are the identity and counter-identity matrices of order 8.

	If the elements of $\mathbf{a}$ are in the set $ \mathcal{C}$,
then $\mathbf{T}_{16}(\mathbf{a})$ is
a low-complexity matrix.
The computational cost of the resulting $2N$-point transform is given by twice the number of bit-shifting operations of the original $N$-point transform; and twice the number of additions plus $2N$ extra additions.

Submitting the optimal 8-point DCT approximations from Table~\ref{T:Avaliacao8} to the JAM method results in fifteen novel 16-point transforms, which are shown in Table~\ref{T:Avaliacao8-16JAM}.
We denote these 16-point DCT approximations
as $\mathbf{\hat C}_{16,j}$, $j = 1, 2, \ldots, 15$.
Table~\ref{T:Comparacao16} lists
competing 16-point DCT approximations \cite{Silveira201644, Silveira201660, Jrid2015, bas2013, bas2010,Oliveira2019, bayer201216pt} for comparison purposes.

		\begin{table}
		\centering
		\caption{
		Results for the novel 16-point DCT approximations.
		} \label{T:Avaliacao8-16JAM}
			\begin{tabular}{ccccccc}
				\hline
				$j$ & $\epsilon(\cdot)$ &	MSE$(\cdot)$ &	$C_g^*(\cdot)$ &	$\eta(\cdot)$ & $\mathcal{A}(\cdot)$ & $\mathcal{S}(\cdot)$\\
				\hline
				1 & 25.13&	0.07&	8.16&	70.98&	\textbf{48}&	\textbf{0} \\
				2 & 24.27&	0.07&	8.16&	70.80&	52&	\textbf{0} \\
				3 & 21.75&	0.07&	8.16&	71.25&	52&	2 \\
				4 & 20.88&	\textbf{0.06}&	8.16&	71.48&	52&	2 \\
				5 & 23.02& 	\textbf{0.06}&	8.37&   71.83&	52&	4 \\
				6 & 22.46&	\textbf{0.06}&	8.18&	71.29&	56&	\textbf{0} \\
				7 & 20.02& 	\textbf{0.06}&	8.16&	71.30&	56&	2 \\
				8 & 20.89&	\textbf{0.06}&   8.16&	71.06&	56&	2 \\
				9 & 18.77&	\textbf{0.06}&	8.37&	72.34&	56&	6 \\
				10 & 19.64&	\textbf{0.06}&	8.37&	72.10&	56&	6 \\
				11 & 18.29&	\textbf{0.06}&	8.19&	70.83&	60&	\textbf{0} \\
				12 & 20.35&	\textbf{0.06}&	8.38&	72.14&	60&	4 \\
				13 & 18.52&	\textbf{0.06}&	8.36&	72.63&	60&	4 \\
				14 & \textbf{16.18}&	\textbf{0.06}&	8.40&	71.67&	64&	4 \\
				15 & 16.41&	\textbf{0.06}&	\textbf{8.57}& 	\textbf{73.51}&	64&	8 \\
				\hline
			\end{tabular}
	\end{table}

	\begin{table}
		\centering
	\setlength{\tabcolsep}{2.4pt}
		\caption{
		Results for competing 16-point DCT approximations.
		} \label{T:Comparacao16}
			\begin{tabular}{lcccccc}
				\hline
				Transform & $\epsilon(\cdot)$ &	MSE$(\cdot)$ &	$C_g^*(\cdot)$ &	$\eta(\cdot)$ & $\mathcal{A}(\cdot)$ & $\mathcal{S}(\cdot)$\\
				\hline
				SOBCM \cite{Silveira201644} & 41.00 & 0.09 & 7.86 & 67.61 & \textbf{44} & \textbf{0} \\
				SBCKMK \cite{Silveira201660} & 30.32 & 0.06 & 8.29 & 70.83 & 60 & \textbf{0} \\
				JAM$_{16}$ \cite{Jrid2015} & 14.74 & \textbf{0.05} & 8.43 & 72.23 & 60 & \textbf{0} \\
				BAS$_{16}$-2013 \cite{bas2013} & 54.62 & 0.13 & 8.19 & 70.64 & 64 & \textbf{0} \\
				BAS$_{16}$-2010 \cite{bas2010} &  16.41 & 0.06 & 8.52 & 73.63 & 64 & 8 \\
				ABM$_{16}$ \cite{Oliveira2019} & 13.70 &  \textbf{0.05} &  \textbf{8.88} & \textbf{76.81} & 64 & 12 \\
				BCEM \cite{bayer201216pt} & \textbf{8.08} & \textbf{0.05} & 7.84 & 65.28 & 72 & \textbf{0} \\
				\hline
			\end{tabular}
	\end{table}

The approximation SOBCM presents the smallest computational cost, requiring 44 additions only.
Although $\mathbf{\hat{C}}_{16,1}$ needs four extra additions,
it outperforms SOBCM, as shown in Tables \ref{T:Avaliacao8-16JAM} and \ref{T:Comparacao16}.
Among the transforms that require 52 additions,
$\mathbf{\hat{C}}_{16,4}$
achieves the smallest total error energy, and
$\mathbf{\hat{C}}_{16,5}$
is the best-performing in terms of MSE, coding gain,
and transform efficiency measures.
Approximation $\mathbf{\hat{C}}_{16,9}$  performs better than competing approaches
that require 56 additions under all considered metrics.
Among the transforms that require 60 additions, JAM$_{16}$ DCT approximation is the best-performing in all evaluated metrics, except transform efficiency, which is maximized for $\mathbf{\hat{C}}_{16,13}$.
The
approximation
ABM$_{16}$
is the best approximation among the considered
64-addition transforms under all discussed metrics.

\subsection{Novel 32-point DCT approximations}

Analogously
to the 16-point case described in the previous section,
we submitted
the discussed
8-point DCT approximation
formalism (Equation~\eqref{E:P8})
to
two instantiations
of the JAM method.
The resulting
$32 \times 32$ DCT approximation matrices
have the following characterization:
\begin{align*}
\mathbf{T}_{32}(\mathbf{a}) =
\begin{bmatrix}
\mathbf{T}_{16}(\mathbf{a}) & \mathbf{0}_{16} \\
\mathbf{0}_{16} & \mathbf{T}_{16}(\mathbf{a})
\end{bmatrix}
\cdot
\begin{bmatrix}
\mathbf{I}_{16} &  \overline{\mathbf{I}}_{16} \\
\mathbf{I}_{16} & -\overline{\mathbf{I}}_{16}
\end{bmatrix}
\end{align*}
where
$\mathbf{0}_{16}$ is $16\times16$
a matrix of zeros;
$\mathbf{I}_{16}$
and
$\overline{\mathbf{I}}_{16}$
are the identity and counter-identity matrices of order 16.

The similarity to the DCT and coding efficiency measurements for all the optimal 8-point DCT approximations scaled to $32\times 32$ are shown in
Table~\ref{T:Avaliacao16-32JAM}, and the 32-point DCT approximations
are denoted by $\mathbf{\hat C}_{32,j}$, $j = 1, 2, \ldots, 15$.
To the best of our knowledge,
the fifteen 32-point transforms listed in Table~\ref{T:Avaliacao16-32JAM}
are new contributions to literature.
For comparison purposes,
Table~\ref{T:Comparacao32}
lists
the performance
of 32-point approximations
found in literature \cite{Jrid2015,bas2013,bas2010,Oliveira2019}.

	\begin{table}
		\centering
		\caption{
		Results for the novel 32-point DCT approximations.}
		\label{T:Avaliacao16-32JAM}
			\begin{tabular}{ccccccc}
				\hline
				$j$ & $\epsilon(\cdot)$ &	MSE$(\cdot)$ &	$C_g^*(\cdot)$ &	$\eta(\cdot)$ & $\mathcal{A}(\cdot)$ & $\mathcal{S}(\cdot)$\\
				\hline
				1  & 68.13&  0.13&   8.23&  56.18&	\textbf{128}&    \textbf{0} \\
				2  & 65.78&  0.13&   8.23& 56.05&	136&	\textbf{0} \\
				3  & 60.57&  0.13&   8.23&  56.43&	136&	4 \\
				4  & 59.47&  0.13&   8.23& 56.78&	136&	4 \\
				5  & 63.93&  \textbf{0.12}&   8.44& 56.72&	136&	8 \\
				6  & 60.69&  \textbf{0.12}&   8.25& 56.47&	144&	\textbf{0} \\
				7  & 57.12&  \textbf{0.12}&   8.23& 56.65&	144&	4 \\
				8  & 58.22&  \textbf{0.12}&   8.23& 56.31&	144&	4 \\
				9  & 55.27&  \textbf{0.12}&   8.44& 57.33&	144&	12 \\
				10 & 56.37&  \textbf{0.12}&   8.44& 56.98&	144&	12 \\
				11 & 52.23&  \textbf{0.12}&   8.27& 56.03&	152&	\textbf{0} \\
				12 & 56.49&  \textbf{0.12}&   8.46& 57.01&	152&	8 \\
				13 & 52.93&  \textbf{0.12}&   8.44& 57.57&	152&	8 \\
				14 & \textbf{48.04}&  \textbf{0.12}&   8.48& 56.57&	160&	8 \\
				15 & 48.73&  \textbf{0.12}&   \textbf{8.65}& \textbf{58.14}&	160&	16 \\
				\hline
			\end{tabular}
	\end{table}

	\begin{table}
		\centering
	\setlength{\tabcolsep}{2.2pt}
		\caption{
		Results for competing 32-point DCT approximations.} \label{T:Comparacao32}
			\begin{tabular}{lcccccc}
				\hline
				Transform & $\epsilon(\cdot)$ &	MSE$(\cdot)$ &	$C_g^*(\cdot)$ &	$\eta(\cdot)$ & $\mathcal{A}(\cdot)$ & $\mathcal{S}(\cdot)$\\
				\hline
				JAM$_{32}$ \cite{Jrid2015}      &  48.10 & \textbf{0.11} & 8.50 & 56.97 & \textbf{152} & \textbf{0} \\
				BAS$_{32}$-2013 \cite{bas2013} & 192.18 & 0.76 & 8.27 & 55.91 & 160 & \textbf{0} \\
				BAS$_{32}$-2010 \cite{bas2010} & 117.07 & 0.24 & 8.50 & 58.50 & 160 & 16 \\
				ABM$_{32}$ \cite{Oliveira2019} &  \textbf{46.27} & \textbf{0.11} & \textbf{8.95} & \textbf{61.03} & 160 & 24 \\
				\hline
			\end{tabular}
	\end{table}

One may
notice
that
the proposed 32-point DCT approximations
demand $15\%$ fewer additions than any previously reported transform.
Among the DCT approximations that require 136 additions, $\mathbf{\hat{C}}_{32,4}$ is the best-performing in terms of total error energy and transform efficiency;
whereas $\mathbf{\hat{C}}_{32,5}$
performed well
in terms of MSE and coding gain.
Similar behavior can be seen for transforms requiring 144 additions.
	Regarding transforms demanding 152 additions,  $\mathbf{\hat{C}}_{32,13}$ outperforms the 32-point
	approximate DCT in~\cite{Jrid2015}
in terms of transform efficiency.
Finally, ABM$_{32}$ is the best-performing among approximations
requiring 160 additions.

\section{Image compression experiments}

\label{ss:imagecoding}

	In order to
	further assess the performance of the best-performing transforms
	we adopted the JPEG-like image compression experiment
detailed in \cite{bas2008,bas2008b,bas2009,bas2010,bas2011,Haweel2001}.
	We present the average results
	for
	45 $ 512 \times 512$ grayscale images taken from the public image database available in~\cite{Base}.
	In this experiment,
the input images are firstly subdivided into blocks of size $ 8 \times 8 $, $ 16 \times 16 $, and $ 32 \times 32 $, depending on the considered
transform size.
Then,
each block is forward transformed through
\eqref{eq:dctfwd},
resulting in a blockwise transformed image.
The transformed blocks are
submitted to a simplified quantization step where the first $r$
coefficients, according to the zig-zag sequence \cite{Wallace1992}, are retained and the remainder are set to zero.
Although the zig-zag sequence could easily be modified to better suit
the considered transforms,
we maintained the standard zig-zag
to facilitate fair comparison with other methods in literature.
Here, $r$ ranges from $ 25 \% $ to $ 99 \% $
the number of coefficients per block.
Finally, the blockwise inverse transform in \eqref{eq:dctbwd} is applied.
In Equations \eqref{eq:dctfwd} and \eqref{eq:dctbwd}
we replaced the transform matrix $\mathbf{C}_N$ by each optimal DCT approximation
$\mathbf{\hat C}_{N,j}$, $j = 1, 2, \ldots, 15$ and $N = 8,16,32$.

From the compressed images, we measure the peak signal-to-noise ratio (PSNR) \cite{Thu2008} and the structural similarity index (SSIM) \cite{Wang2004}.
	The former is a traditional and widely used
	metric
	for image quality assessment \cite{Thu2008}.
	The latter is a complementary
	figure of merit
	for evaluating
	image quality
	that considers luminance, contrast, and structure of the image to quantify
	degradation,
	approaching to a subjective
	assessment
	\cite{Wang2011}.
    Besides presenting the average PSNR and SSIM curves, for better visualization, we show their absolute percentage error (APE) \cite{Hig2008}.
The presented APE results consider the DCT-based measurements as baseline.

\begin{figure*}
		\centering
			\subfigure[PSNR.]{
				{\includegraphics[width=0.45\textwidth]{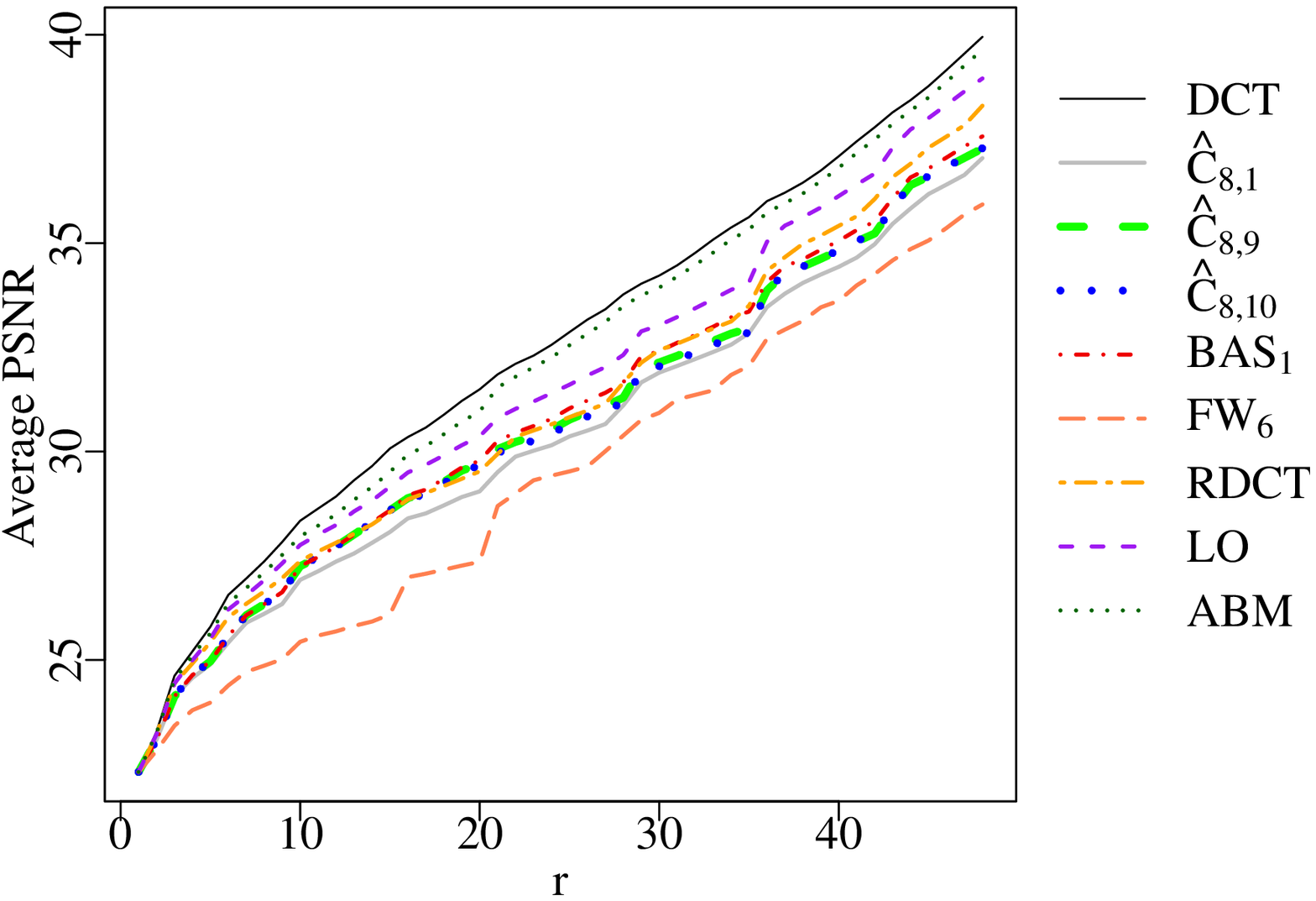}}\label{F:PSNR8}}
			\subfigure[SSIM.]{
				{\includegraphics[width=0.45\textwidth]{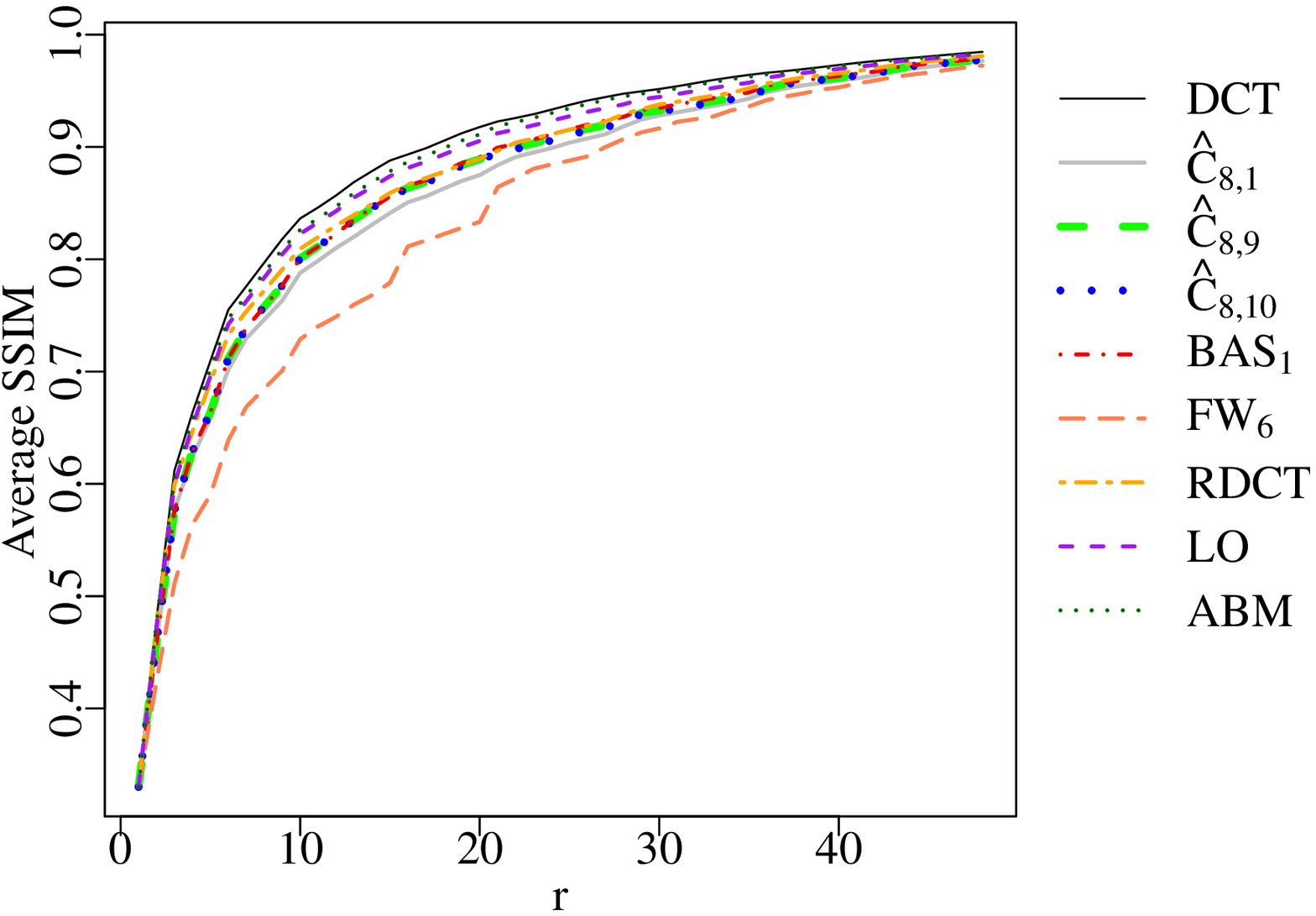}}\label{F:MSSIM8}}
			\subfigure[APE (PSNR).]{
				{\includegraphics[width=0.45\textwidth]{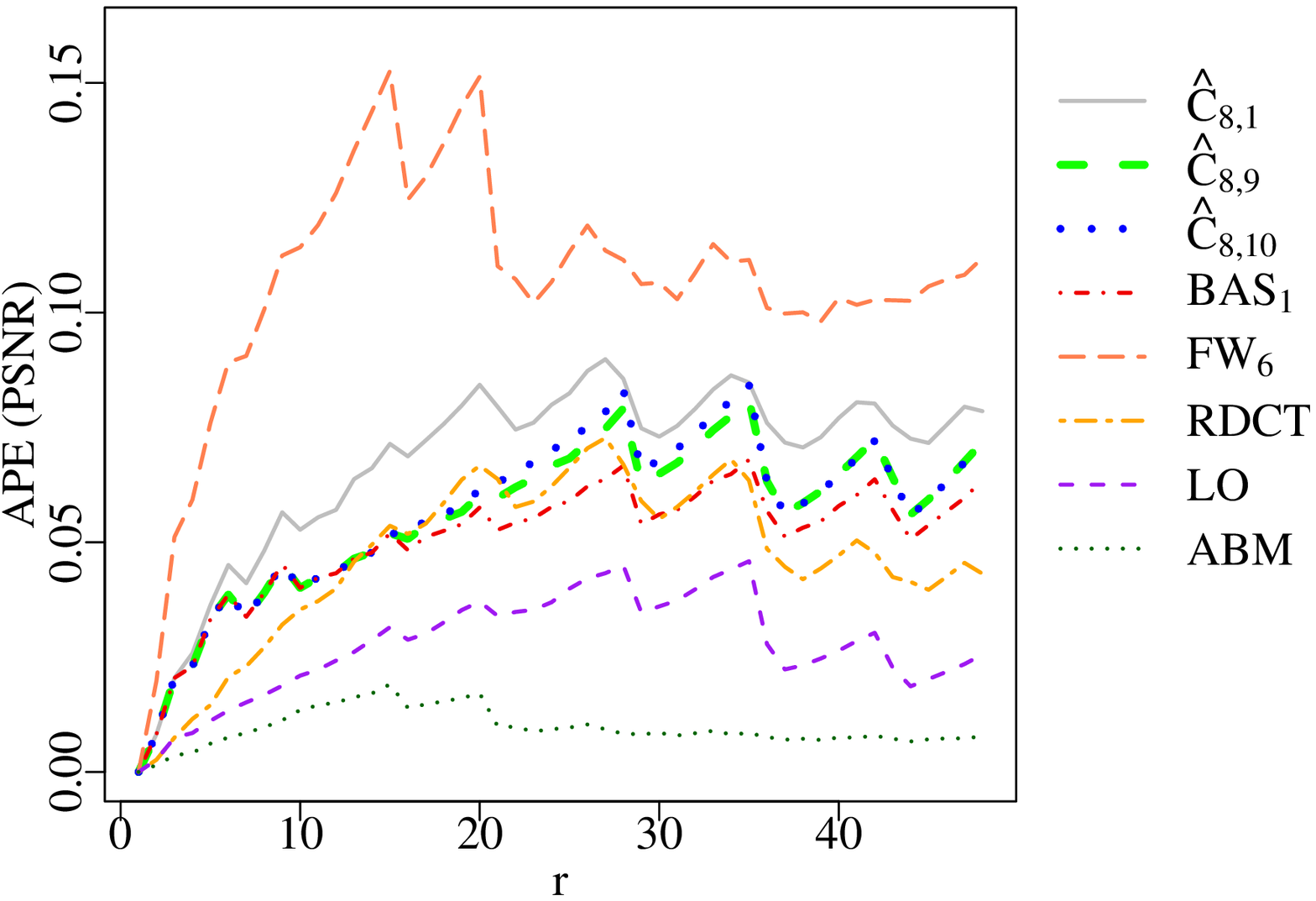}}\label{F:ape.PSNR8}}
			\subfigure[APE (SSIM).]{
				{\includegraphics[width=0.45\textwidth]{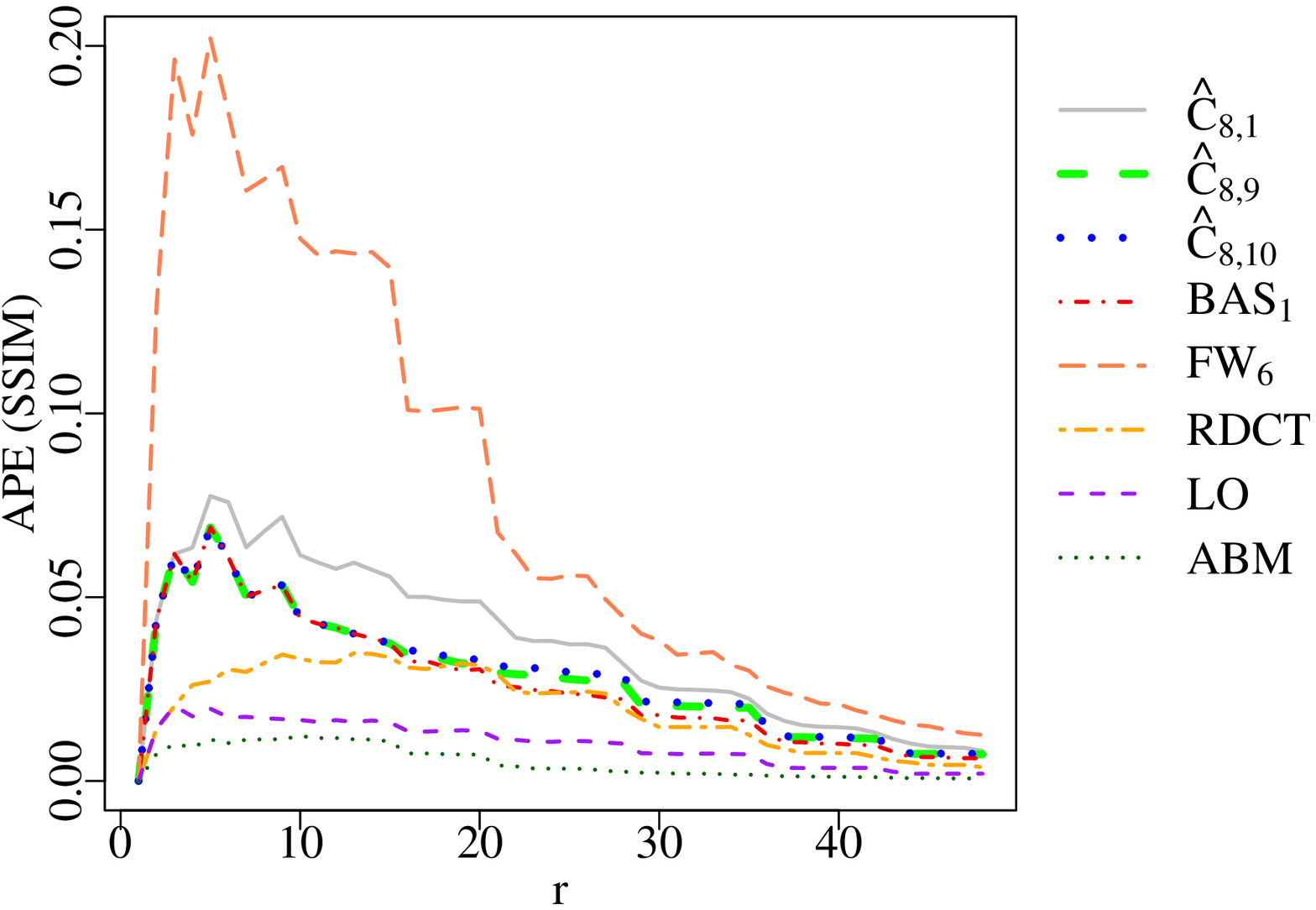}}\label{F:ape.MSSIM8}}
			\subfigure[PSNR$/\mathcal{A}$.]{
				{\includegraphics[width=0.45\textwidth]{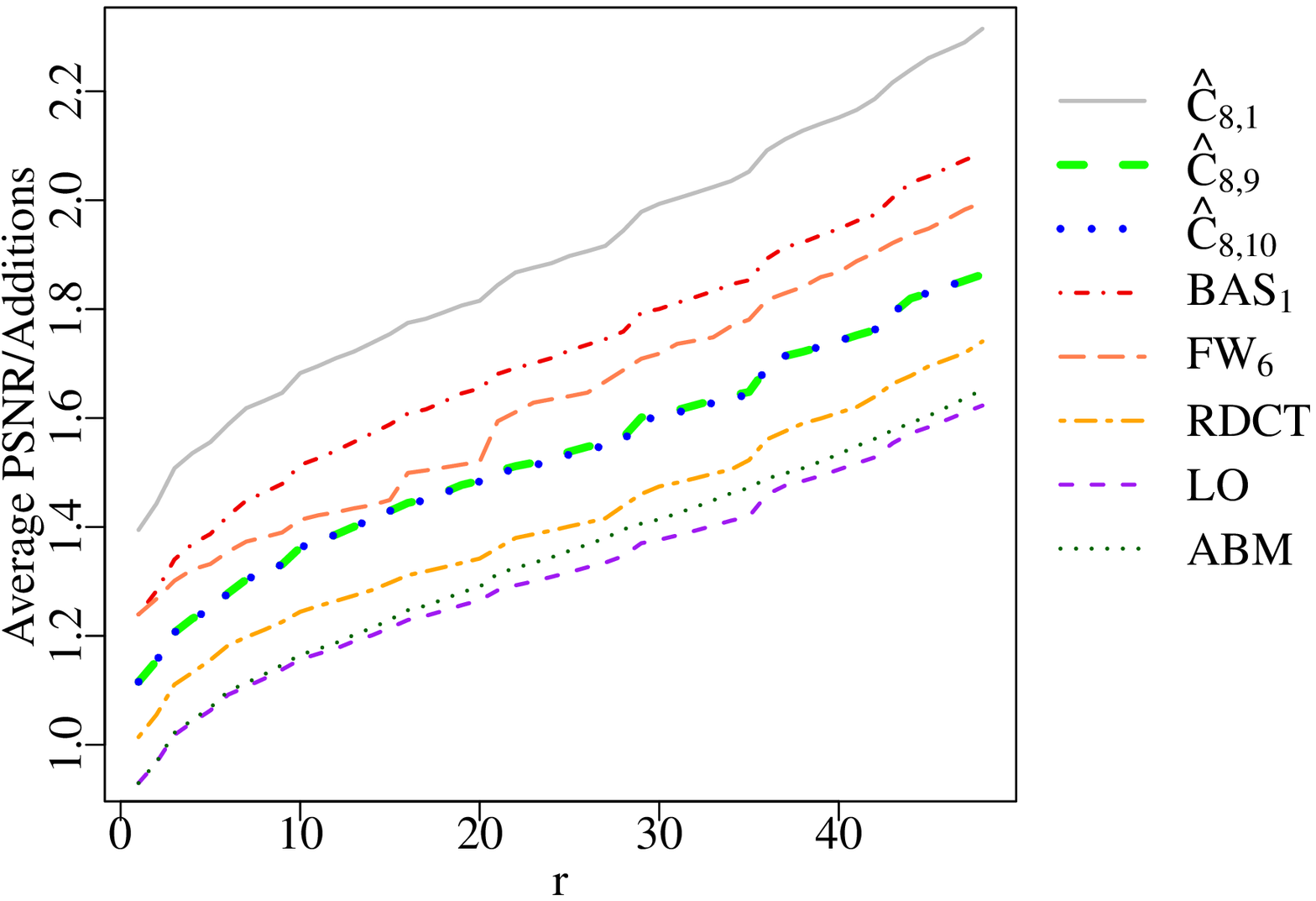}}\label{F:ape.PSNR.add8}}
			\subfigure[SSIM$/\mathcal{A}$.]{
				{\includegraphics[width=0.45\textwidth]{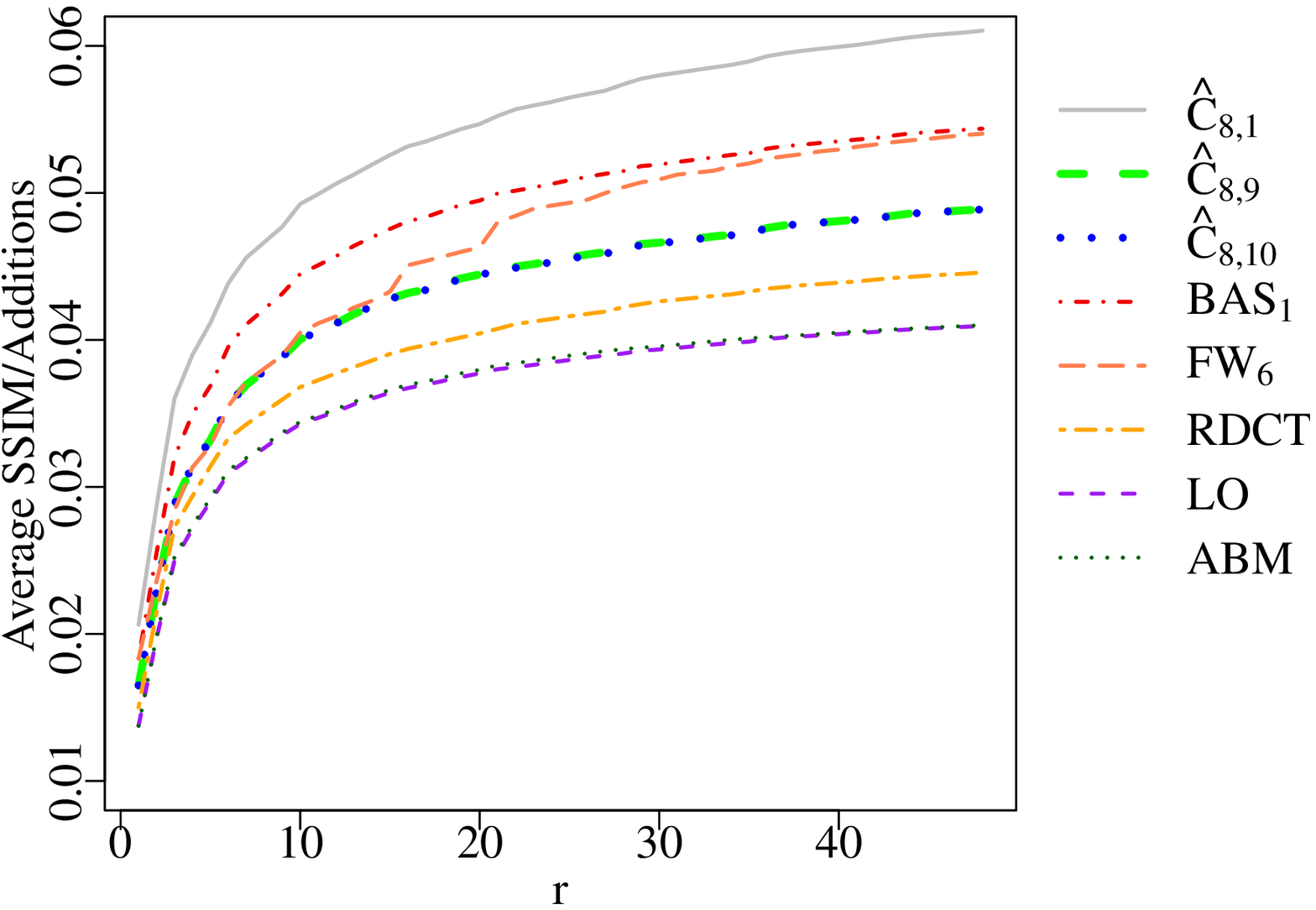}}\label{F:ape.MSSIM.add8}}
			\caption{
			Image compression results for novel and competing 8-point DCT approximations.
			}
			\label{F:compcompre8}
	\end{figure*}

For the 8-point transforms,
Figures~\ref{F:PSNR8}, \ref{F:MSSIM8}, \ref{F:ape.PSNR8} and \ref{F:ape.MSSIM8}
show that
the ABM approximation
	performs the best.
Considering transforms that require 20 addition,
$\hat{\mathbf{C}}_{8,9}$ offers the best
results and FW$_6$ performs unfavorably.
These results corroborate the coding gain measurements presented in Tables \ref{T:Avaliacao8} and \ref{T:Comparacao8}.
Additionally, Figures~\ref{F:ape.PSNR.add8} and \ref{F:ape.MSSIM.add8} show the gain in terms of PSNR and SSIM per addition operation.
The approximation $\hat{\mathbf{C}}_{8,1}$ offered the highest PSNR and SSIM gain per addition.
	A median behavior is achieved by $\hat{\mathbf{C}}_{8,9}$ in terms of PSNR, SSIM, and the gains by addition.
The only transforms requiring 20~additions are found in~\cite{Tablada2015}.
Thus the approximation $\hat{\mathbf{C}}_{8,9}$ is the best one at 20 additions.

\begin{figure*}
		\centering
			\subfigure[PSNR.]{
				{\includegraphics[width=0.45\textwidth]{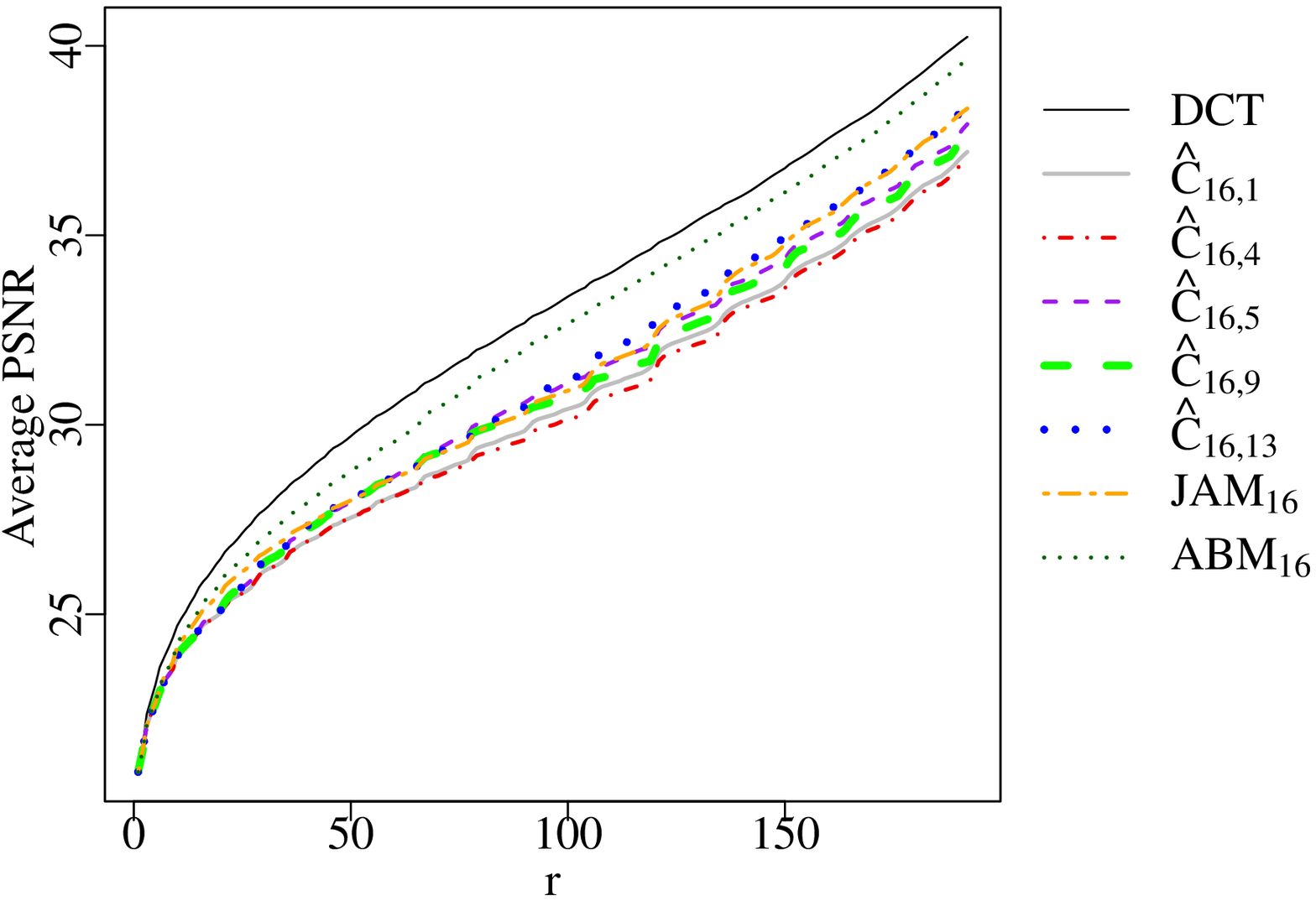}}\label{F:PSNR16}}
			\subfigure[SSIM.]{
				{\includegraphics[width=0.45\textwidth]{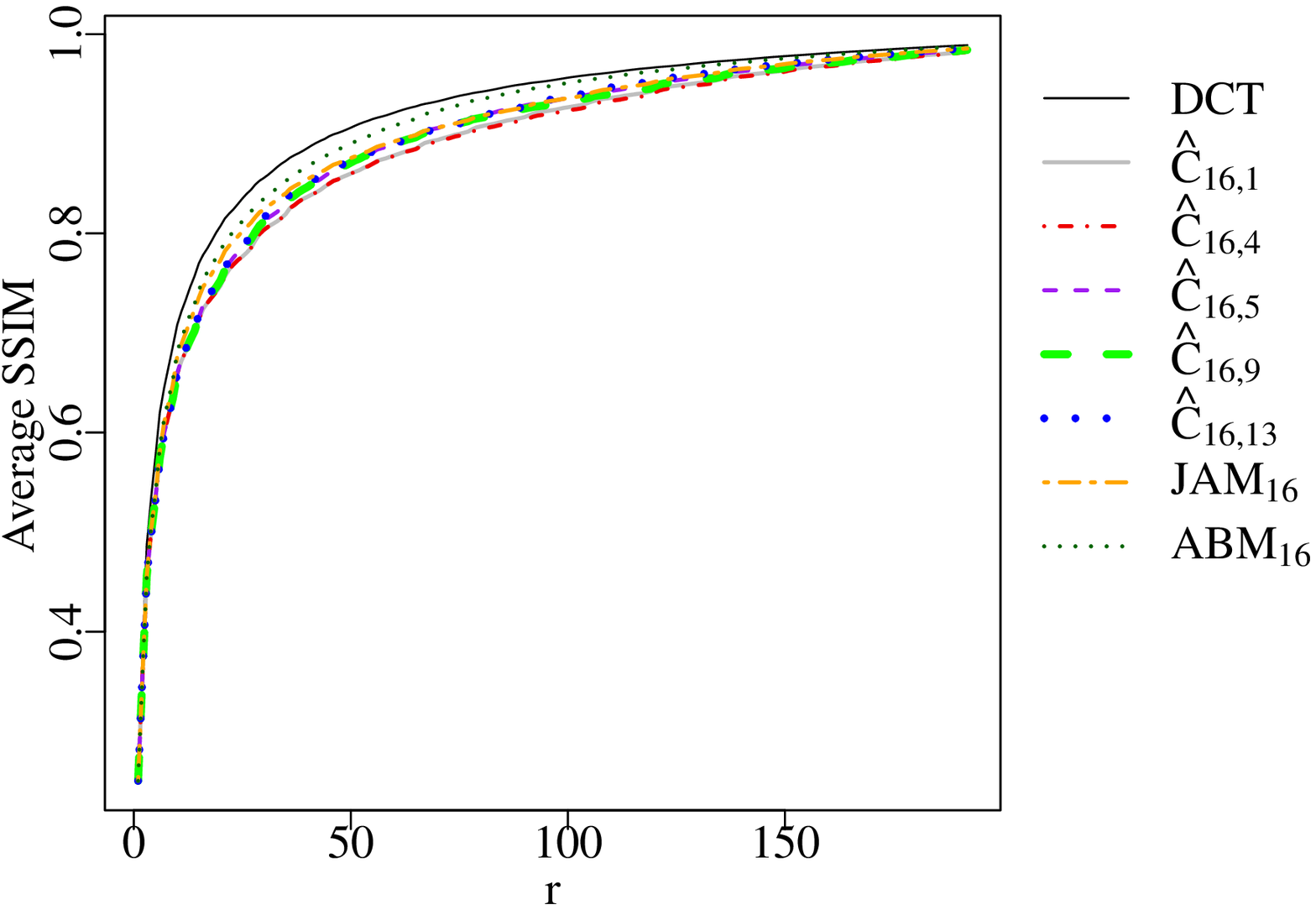}}\label{F:MSSIM16}}
			\subfigure[APE (PSNR).]{
				{\includegraphics[width=0.45\textwidth]{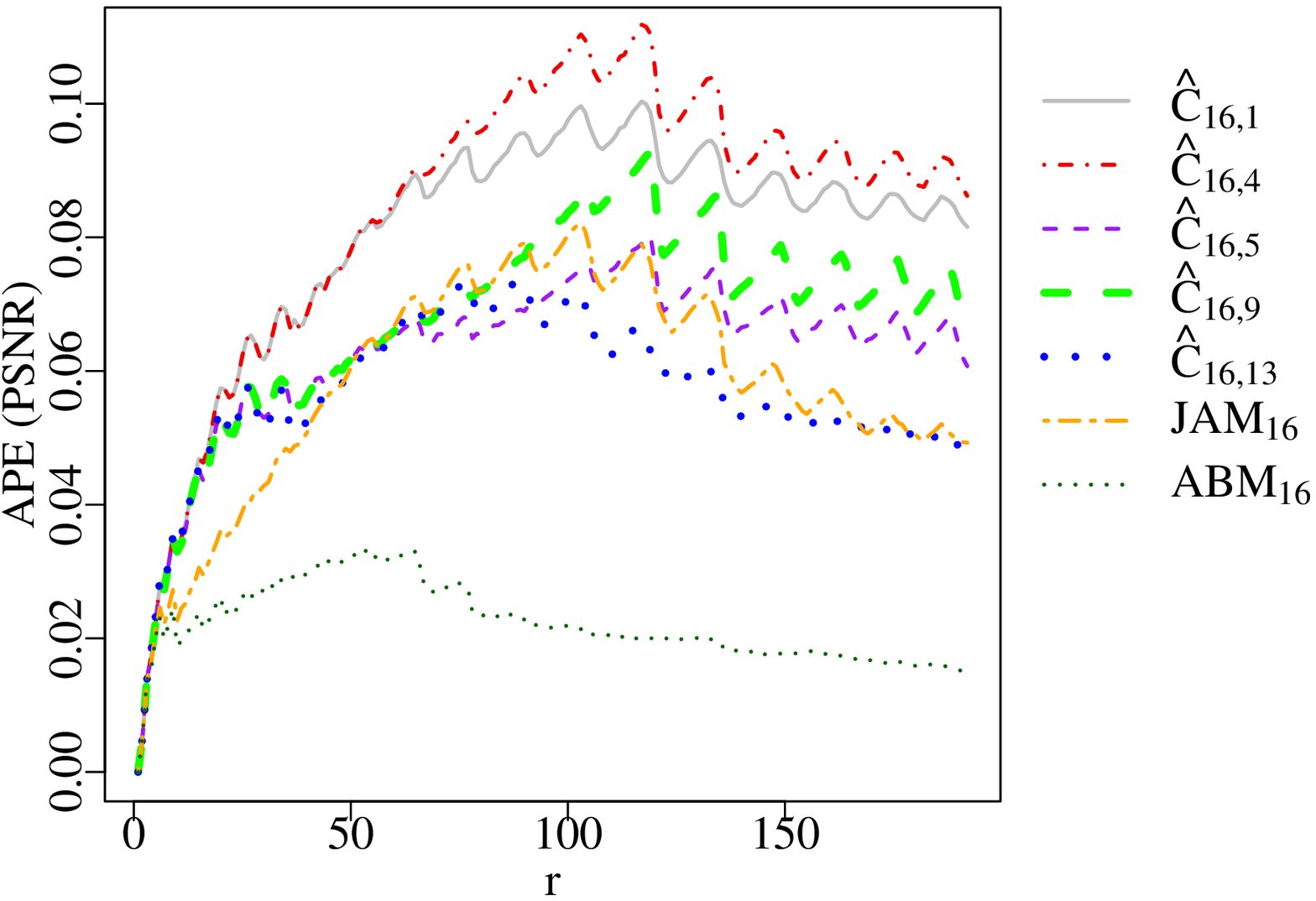}}\label{F:ape.PSNR16}}
			\subfigure[APE (SSIM).]{
				{\includegraphics[width=0.45\textwidth]{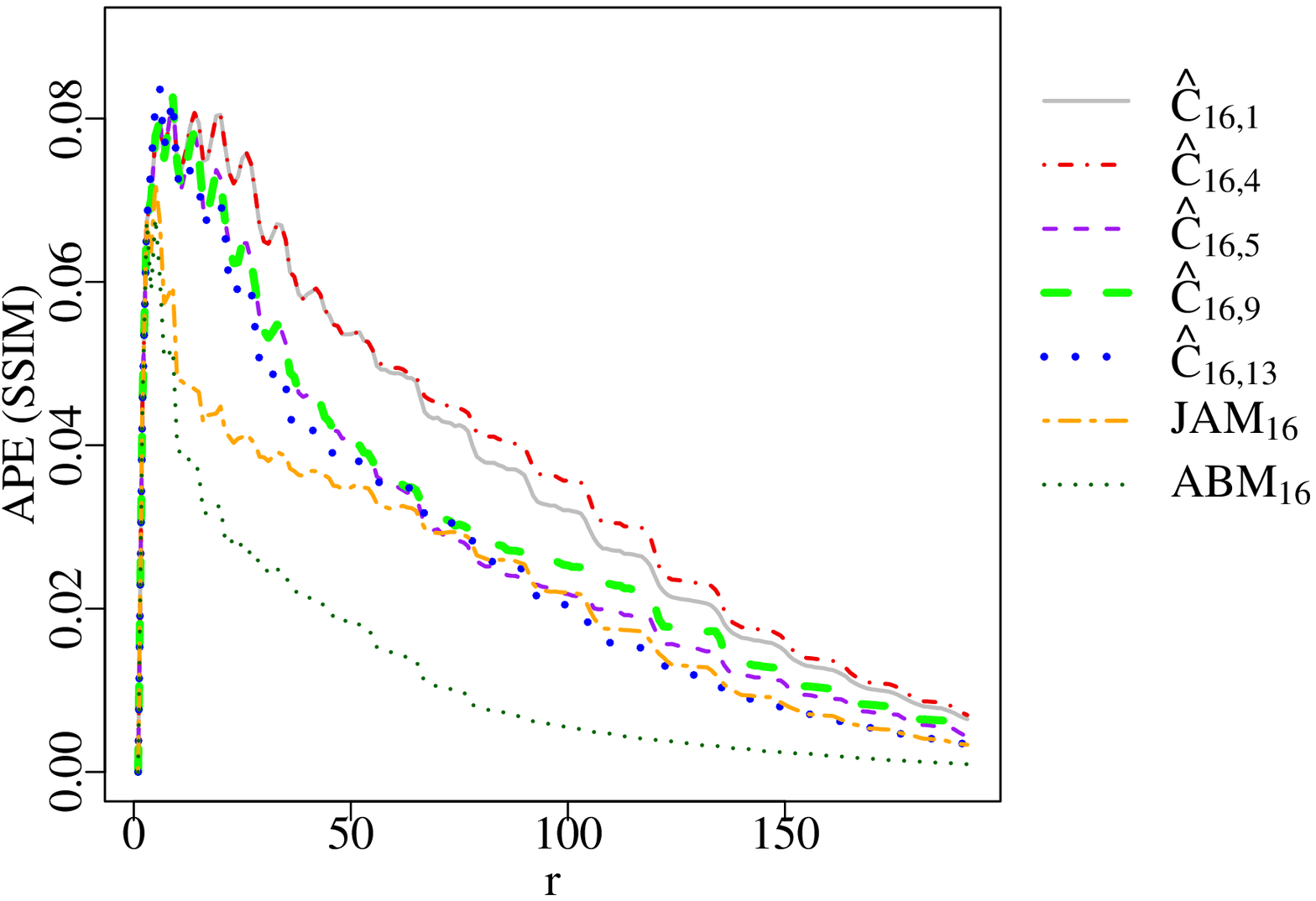}}\label{F:ape.MSSIM16}}
			\subfigure[PSNR$/\mathcal{A}$.]{
				{\includegraphics[width=0.45\textwidth]{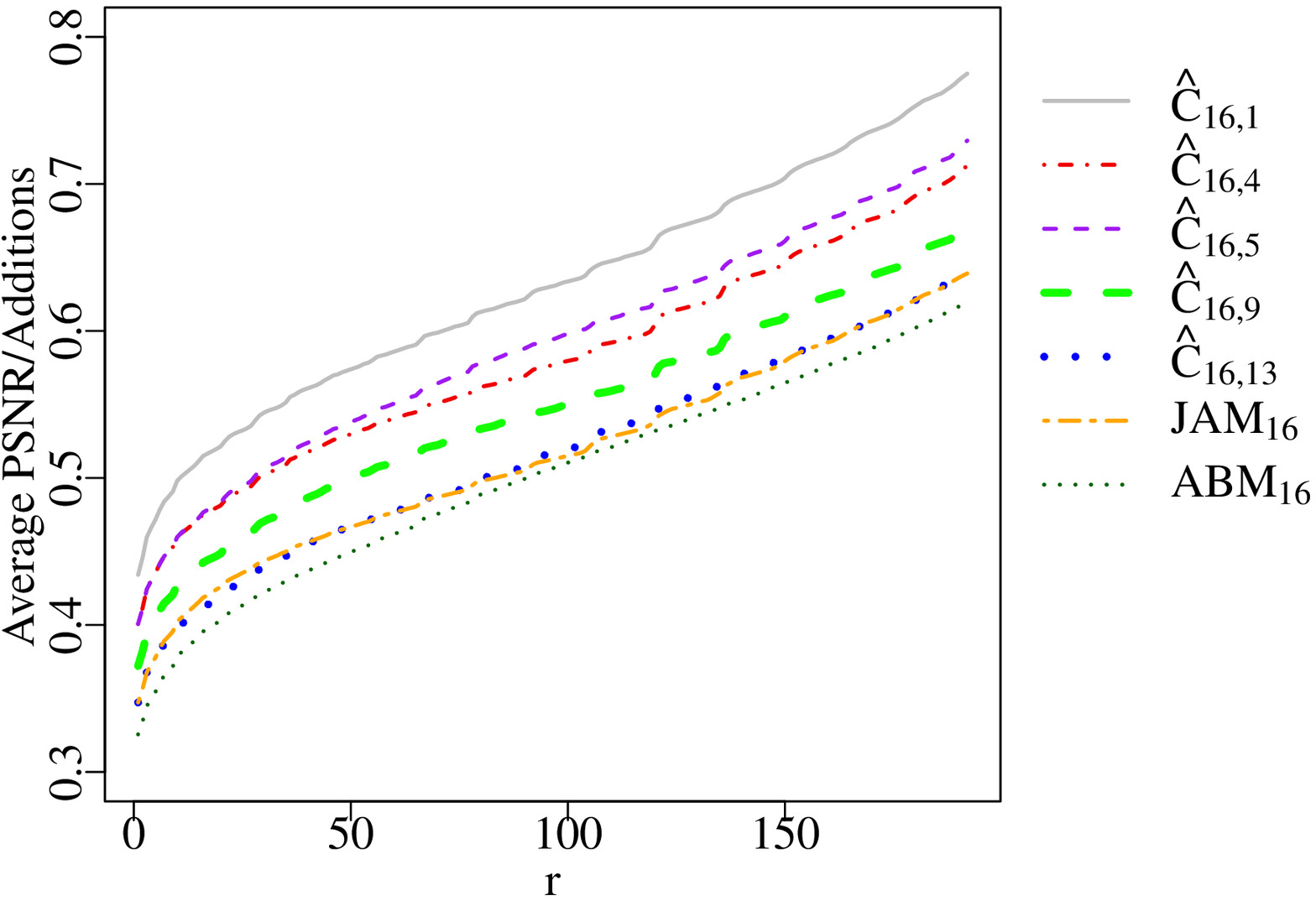}}\label{F:ape.PSNR.add16}}
			\subfigure[SSIM$/\mathcal{A}$.]{
				{\includegraphics[width=0.45\textwidth]{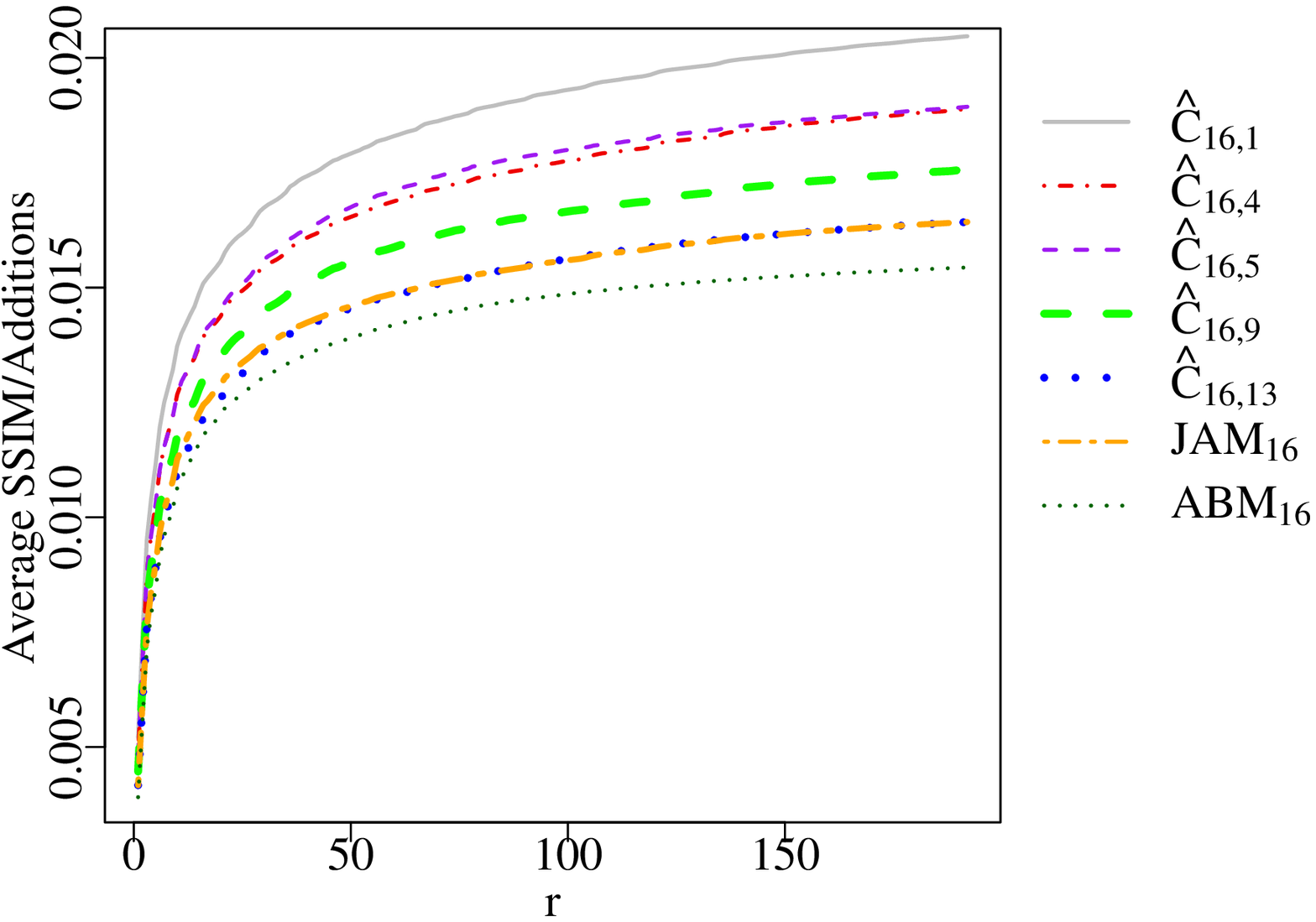}}\label{F:ape.MSSIM.add16}}
			\caption{
			Image compression results for novel and competing 16-point DCT approximations.
			}
			\label{F:compcompre16}
	\end{figure*}

	Considering 16-point DCT approximations,
	according to Figures~\ref{F:PSNR16}, \ref{F:MSSIM16}, \ref{F:ape.PSNR16} and \ref{F:ape.MSSIM16},
	the
approximation
	ABM$_{16}$
	was the best-performing.
Approximation	$\hat{\mathbf{C}}_{16,1}$ has the highest PSNR and SSIM gain per additive cost unit, as shown in Figures~\ref{F:ape.PSNR.add16} and \ref{F:ape.MSSIM.add16}.
	As for the case of 8-point transforms, $\hat{\mathbf{C}}_{16,9}$ has a median behavior for all the considered metrics in this image compression experiment.
In this case, to the best of our knowledge,
there are not 48-, 52- or 56-addition methods for a direct comparison.
	Threfore the proposed transforms, $\hat{\mathbf{C}}_{16,1}$, $\hat{\mathbf{C}}_{16,4}$,
	$\hat{\mathbf{C}}_{16,5}$, and $\hat{\mathbf{C}}_{16,9}$ stand alone as best in their classes.

		\begin{figure*}
		\centering
			\subfigure[PSNR.]{
				{\includegraphics[width=0.45\textwidth]{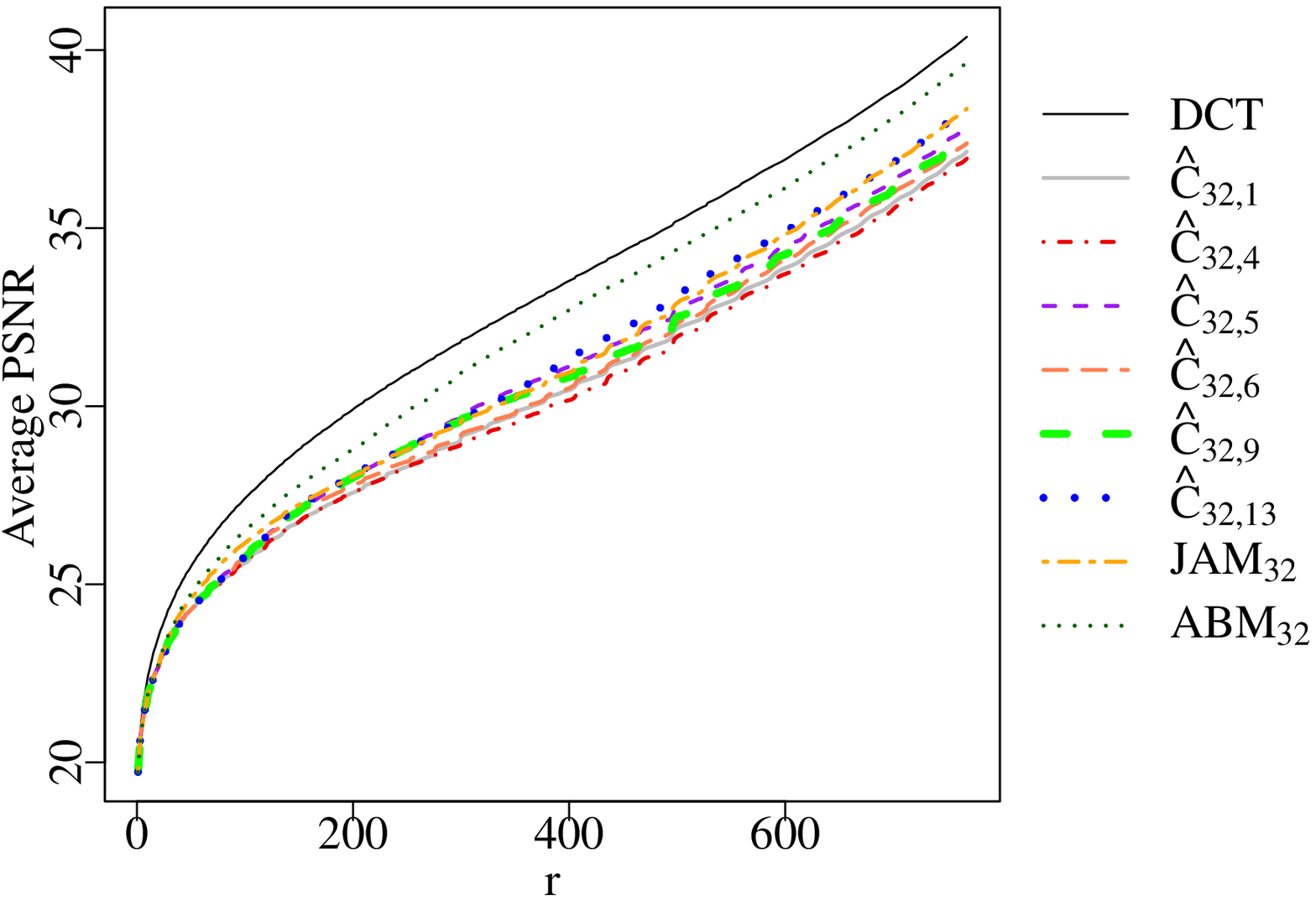}}\label{F:PSNR32}}
			\subfigure[SSIM.]{
				{\includegraphics[width=0.45\textwidth]{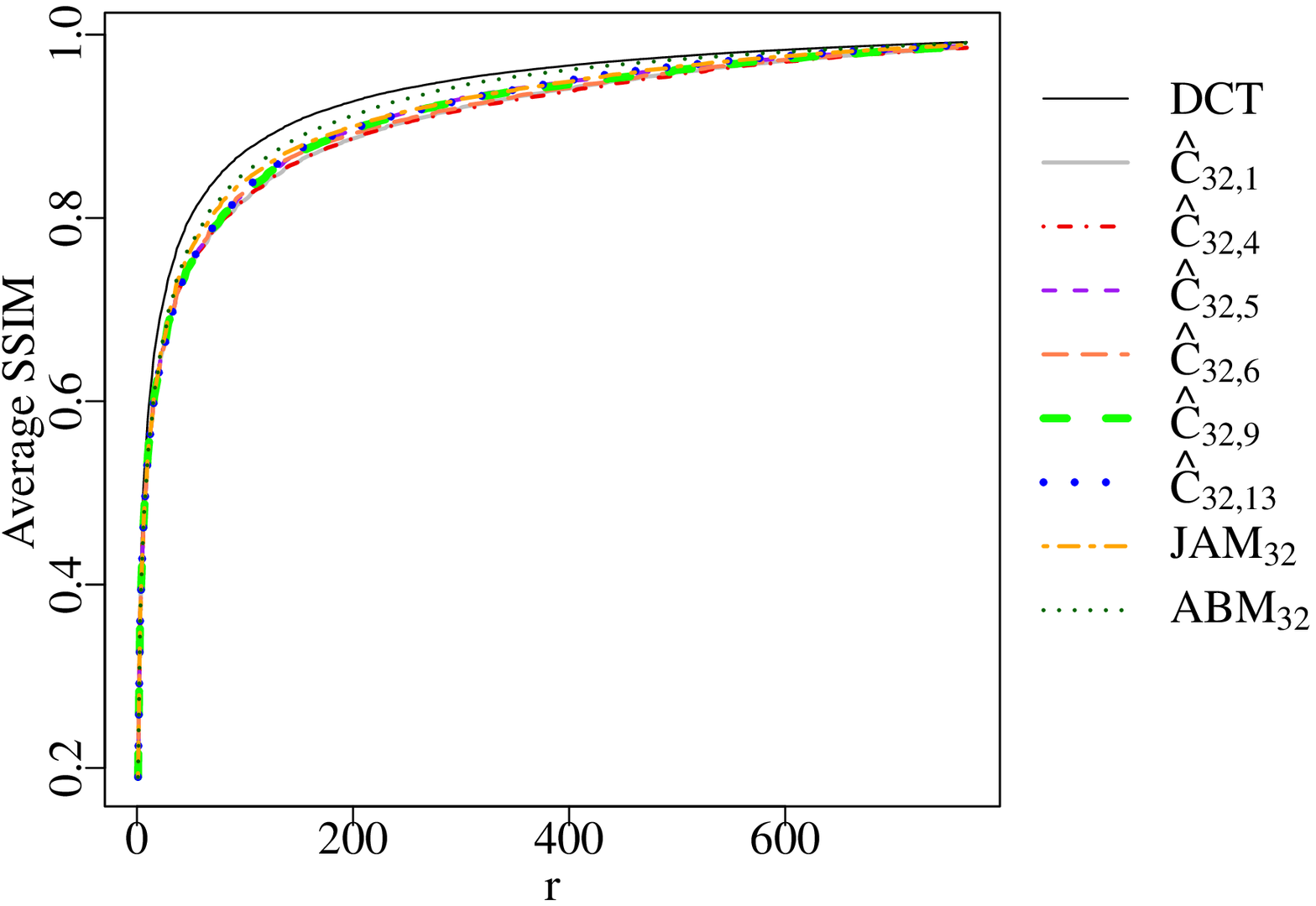}}\label{F:MSSIM32}}
			\subfigure[APE (PSNR).]{
				{\includegraphics[width=0.45\textwidth]{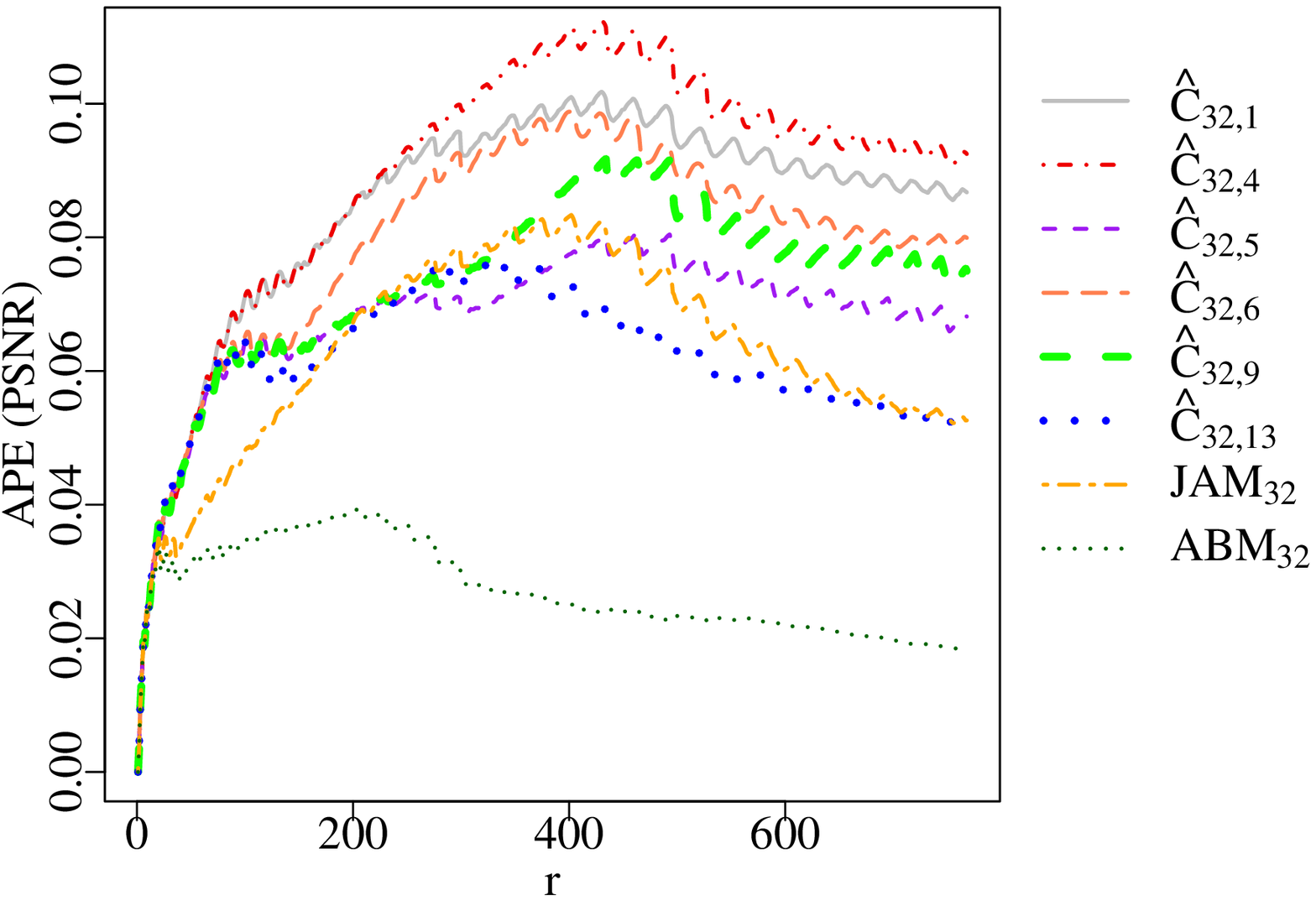}}\label{F:ape.PSNR32}}
			\subfigure[APE (SSIM).]{
				{\includegraphics[width=0.45\textwidth]{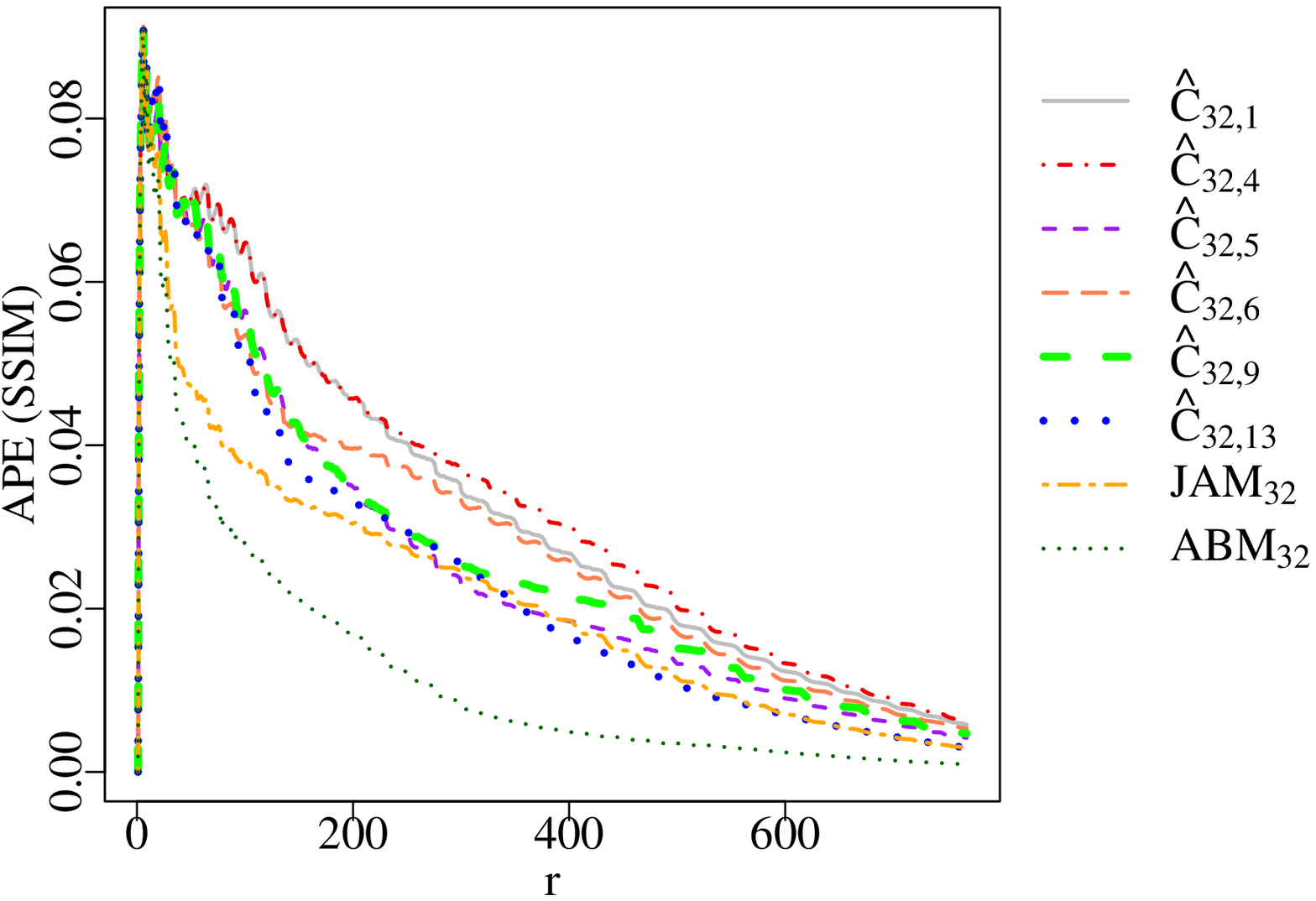}}\label{F:ape.MSSIM32}}
			\subfigure[PSNR$/\mathcal{A}$.]{
				{\includegraphics[width=0.45\textwidth]{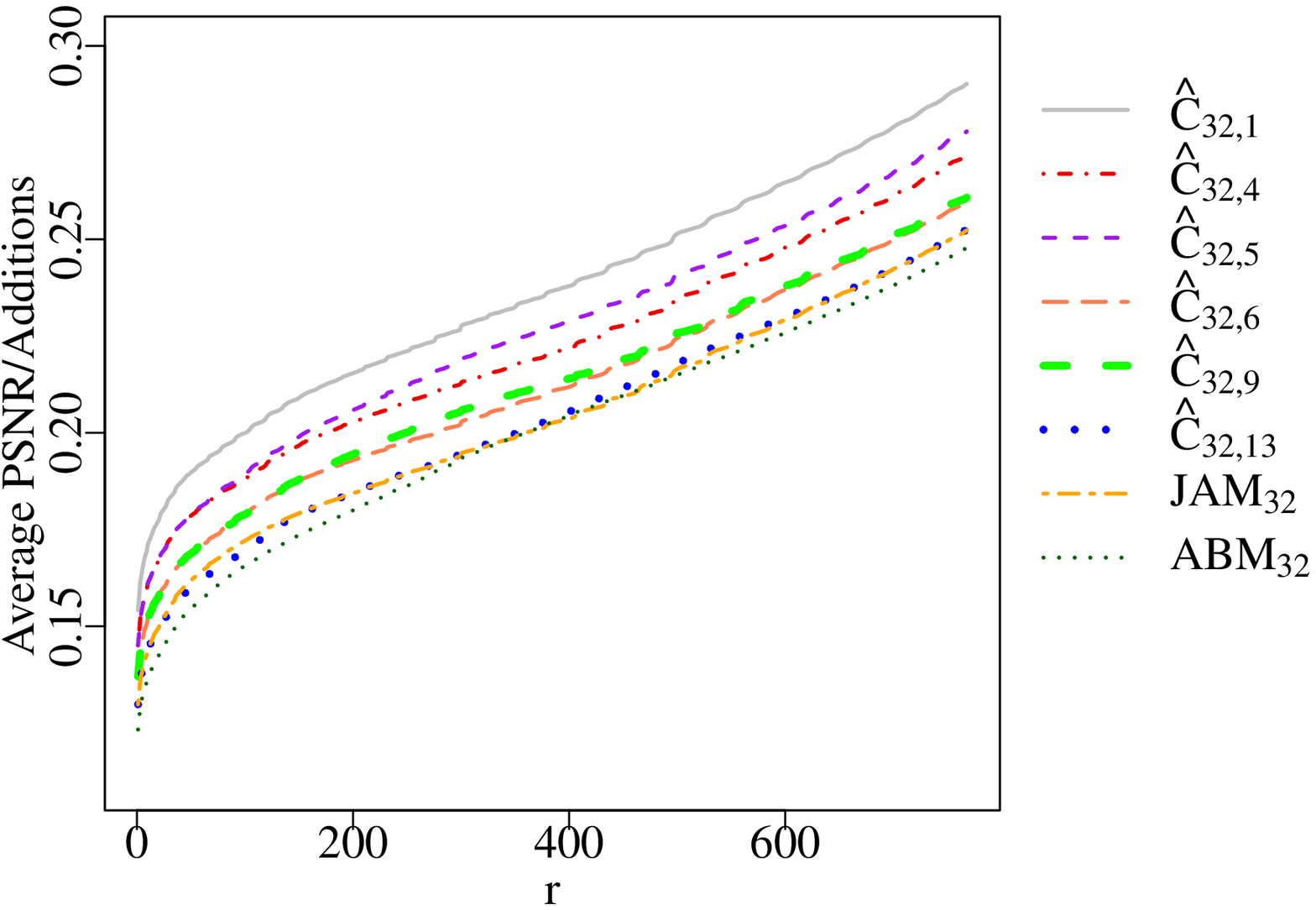}}\label{F:ape.PSNR.add32}}
			\subfigure[SSIM$/\mathcal{A}$.]{
				{\includegraphics[width=0.45\textwidth]{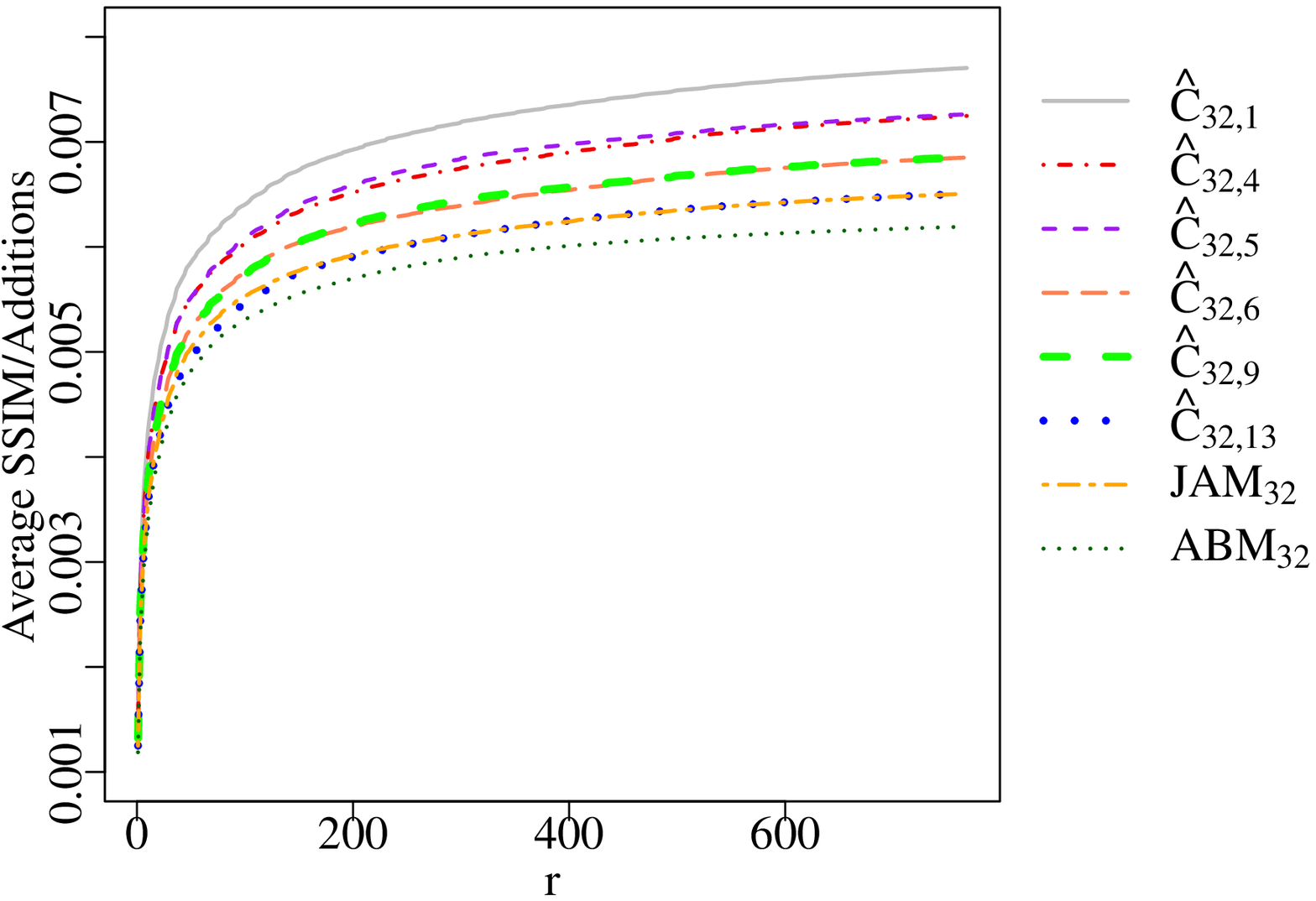}}\label{F:ape.MSSIM.add32}}
			\caption{
			Image compression results for novel and competing 32-point DCT approximations.
			}
			\label{F:compcompre32}
	\end{figure*}

	Transforms of sizes $16\times16$ and $32\times32$ behave very similarly, as shown in Figures \ref{F:compcompre16} and \ref{F:compcompre32} for the 32-point case.
Approximations
$\hat{\mathbf{C}}_{32,1}$, $\hat{\mathbf{C}}_{32,4}$, $\hat{\mathbf{C}}_{32,5}$, $\hat{\mathbf{C}}_{32,6}$, and $\hat{\mathbf{C}}_{32,9}$ have lower arithmetic complexity than
any
already known 32-point DCT approximations in literature.
	Transform $\hat{\mathbf{C}}_{32,1}$ has the
	smaller computational
	complexity.
	For a range of $r$ values the transform $\hat{\mathbf{C}}_{32,13}$ is outperformed by JAM$_{32}$.
	As far as we know, a 32-point transform requiring less than 152 additions is absent in literature.
As a consequence, transforms
$\hat{\mathbf{C}}_{32,1}$, $\hat{\mathbf{C}}_{32,4}$, $\hat{\mathbf{C}}_{32,5}$,
$\hat{\mathbf{C}}_{32,6}$, and $\hat{\mathbf{C}}_{32,9}$ are identified as best performing tools.
The main behavior of the optimal BAS-based 8-point transforms is preserved when scaling them to transforms of size $16 \times 16$ and $32 \times 32$ using the JAM algorithm.

\section{Video coding experiments}

\label{section-video}

In this section, we show the applicability of the best-performing multiparametric DCT
approximations---$\mathbf{\hat{C}}_{8,1}$,
$\mathbf{\hat{C}}_{8,5}$, and $\mathbf{\hat{C}}_{8,9}$ and their 16- and 32-point scaled versions---in the context
of video coding.
We embedded the selected transforms into the public available HEVC reference software
provided in~\cite{refsoft},
then we assessed the performance of the resulting systems.
The HEVC improves its predecessors~\cite{Sullivan2012} and aims at providing high video compression rates~\cite{hevc1}.
Unlike H.264 and older video coding standards, the HEVC standard employs not only an 8-point DCT but also transforms of length 4, 16, and 32 for better handling smooth and textured image regions of various sizes~\cite{hevc1}.

In our experiments, we replace the original 8-, 16-, and 32-point HEVC transforms
by the three selected DCT approximations and their scaled versions, one at a time.
We shall call $\mathbf{\hat{C}}_1$, $\mathbf{\hat{C}}_5$, and $\mathbf{\hat{C}}_9$
each set of transforms based on $\mathbf{\hat{C}}_{8,1}$, $\mathbf{\hat{C}}_{8,5}$, and $\mathbf{\hat{C}}_{8,9}$, respectivelly.
The original 4-point HEVC transform was kept unchanged because it is already a low-complexity transformation.
We encoded the first 100 frames of one video sequence of each A to F class following the recommendations in the Common Test Conditions (CTC) documentation~\cite{CTConditions2013}.
Namely we used the following 8-bit videos:
``PeopleOnStreet'' (2560$\times$1600 at 30~fps),
``BasketballDrive'' (1920$\times$1080 at 50~fps),
``RaceHorses'' (832$\times$480 at 30~fps),
``BlowingBubbles'' (416$\times$240  at 50~fps),
``KristenAndSara'' (1280$\times$720  at 60~fps),
and
``BasketbalDrillText'' (832$\times$480  at 50~fps).
All the encoding parameters were also set according to CTC documentation for the Main profile and All-Intra (AI), Random Access (RA), Low Delay B (LD-B), and Low Delay P (LD-P) configurations.

For assessing image quality, we use the  per color channel MSE (MSE-Y, MSE-U, and MSE-V) and PSNR (PSNR-Y, PSNR-U, and PSNR-V), for each video frame~\cite{Ohm2012}.
Such measurements are collected by the reference software.
From those values, we also calculate the Bj{\o}ntegaard's delta PSNR (BD-PSNR) and delta rate (BD-Rate)~\cite{Bjontegaard2001, Hanhart2014} for all  discussed transform sets and configurations.
The BD-PSNR and DB-Rate measurements for each video sequence are summarized in Table~\ref{tab:bdpsnrbdrate}.

The approximation $\mathbf{\hat{C}}_5$ performed
slightly better performance than $\mathbf{\hat{C}}_9$ in most cases.
Replacing the original HEVC transforms by either $\mathbf{\hat{C}}_5$ or $\mathbf{\hat{C}}_9$ results in a loss of no more than 0.60~dB at AI configuration.
Finally, note that both transforms $\mathbf{\hat{C}}_5$ and $\mathbf{\hat{C}}_9$ have consistently better results than $\mathbf{\hat{C}}_1$ in terms of BD-PSNR and BD-Rate.
These findings reinforce the results from
the still image compression experiments
in the previous section.

\begin{table}
\centering
\caption{Average Bj{\o}ntegaard's metrics for the modified versions of the HEVC reference software.}
\label{tab:bdpsnrbdrate}
\begin{tabular}{cccccccc}
\hline
\multirow{2}{*}{ Config. } & \multirow{2}{*}{ Video sequence } & \multicolumn{3}{c}{ BD-PSNR (dB) } & \multicolumn{3}{c}{ BD-Rate (\%) } \\
\cline{3-8}
& & $\mathbf{\hat{C}}_1$ & $\mathbf{\hat{C}}_5$ & $\mathbf{\hat{C}}_9$ & $\mathbf{\hat{C}}_1$ & $\mathbf{\hat{C}}_5$ & $\mathbf{\hat{C}}_9$  \\
\hline
\multirow{6}{*}{ AI }
& ``PeopleOnStreet'' & $-0.495$ &  $-0.465$ &  $-0.465$ &  $9.870$  &  $9.251$  &  $9.246$  \\
& ``BasketballDrive'' & $-0.267$ &  $-0.250$ &  $-0.253$ &  $10.474$  &  $9.771$  &  $9.870$  \\
& ``RaceHorses'' & $-0.622$ &  $-0.609$ &  $-0.596$ &  $8.135$  &  $7.970$  &  $7.795$  \\
& ``BlowingBubbles'' & $-0.219$ &  $-0.203$ &  $-0.203$ &  $3.880$  &  $3.601$  &  $3.600$  \\
& ``KristenAndSara'' & $-0.415$ &  $-0.389$ &  $-0.392$ &  $8.628$  &  $8.100$  &  $8.143$  \\
& ``BasketballDrillText'' & $-0.172$ &  $-0.162$ &  $-0.162$ &  $3.358$  &  $3.160$  &  $3.168$ \\
\hline
\multirow{5}{*}{ RA }
& ``PeopleOnStreet'' & $-0.264$ &  $-0.244$ &  $-0.245$ &  $6.484$  &  $5.985$  &  $6.009$  \\
& ``BasketballDrive'' & $-0.214$ &  $-0.199$ &  $-0.203$ &  $10.060$  &  $9.330$  &  $9.547$  \\
& ``RaceHorses'' & $-0.816$ &  $-0.780$ &  $-0.746$ &  $13.857$  &  $13.264$  &  $12.762$  \\
& ``BlowingBubbles'' & $-0.158$ &  $-0.144$ &  $-0.148$ &  $4.296$  &  $3.891$  &  $4.014$  \\
& ``BasketballDrillText'' & $-0.223$ &  $-0.207$ &  $-0.207$ &  $5.573$  &  $5.151$  &  $5.153$  \\
\hline
\multirow{5}{*}{ LDB }
& ``BasketballDrive'' & $-0.201$ &  $-0.187$ &  $-0.190$ &  $8.899$  &  $8.242$  &  $8.395$  \\
& ``RaceHorses'' & $-0.805$ &  $-0.771$ &  $-0.739$ &  $12.674$  &  $12.138$  &  $11.678$  \\
& ``BlowingBubbles'' & $-0.160$ &  $-0.151$ &  $-0.153$ &  $4.442$  &  $4.193$  &  $4.243$  \\
& ``KristenAndSara'' & $-0.200$ &  $-0.180$ &  $-0.187$ &  $7.081$  &  $6.319$  &  $6.525$  \\
& ``BasketballDrillText'' & $-0.267$ &  $-0.243$ &  $-0.251$ &  $7.029$  &  $6.382$  &  $6.580$  \\
\hline
\multirow{5}{*}{ LDP }
& ``BasketballDrive'' & $-0.203$ &  $-0.187$ &  $-0.192$ &  $8.945$  &  $8.173$  &  $8.459$  \\
& ``RaceHorses'' & $-0.775$ &  $-0.744$ &  $-0.715$ &  $12.094$  &  $11.607$  &  $11.203$  \\
& ``BlowingBubbles'' & $-0.150$ &  $-0.138$ &  $-0.139$ &  $4.267$  &  $3.906$  &  $3.964$  \\
& ``KristenAndSara'' & $-0.179$ &  $-0.163$ &  $-0.168$ &  $6.668$  &  $6.016$  &  $6.244$  \\
& ``BasketballDrillText'' & $-0.246$ &  $-0.227$ &  $-0.234$ &  $6.565$  &  $6.057$  &  $6.222$  \\
\hline
\end{tabular}
\end{table}

\begin{figure}
\centering
\subfigure[default HEVC DCT]{\includegraphics[scale=.54]{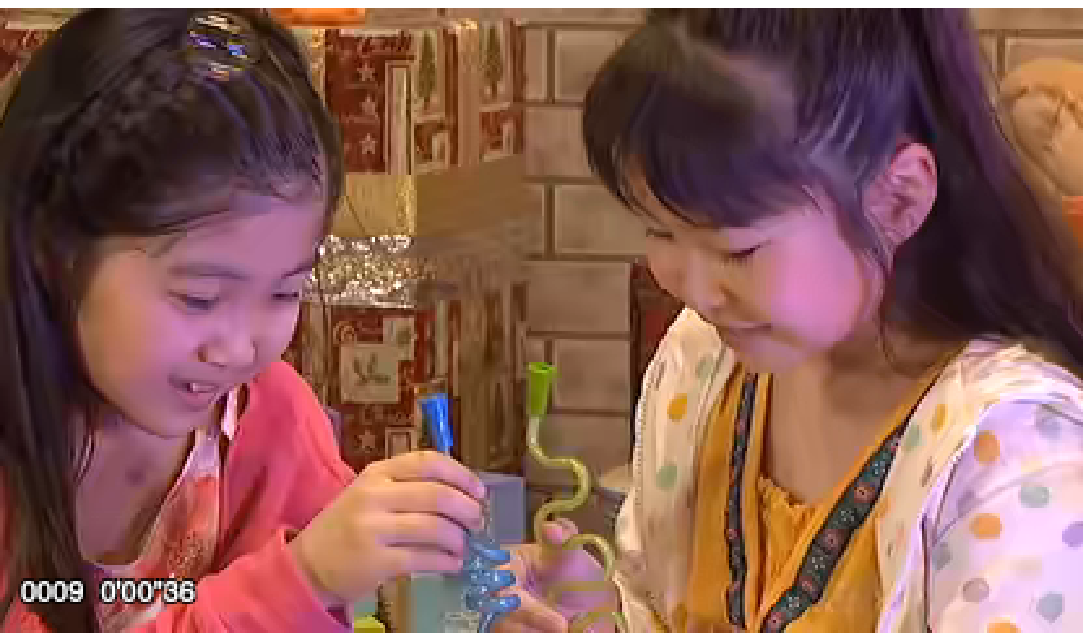}}
\subfigure[$\mathbf{\hat{C}}_1$]{\includegraphics[scale=.54]{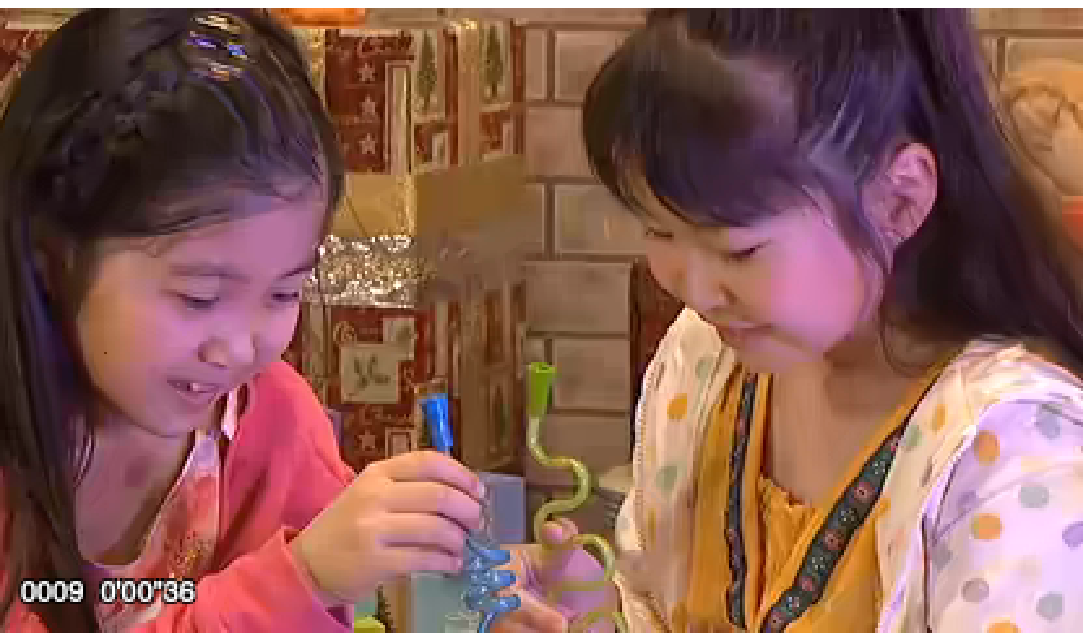}}\\
\subfigure[$\mathbf{\hat{C}}_5$]{\includegraphics[scale=.54]{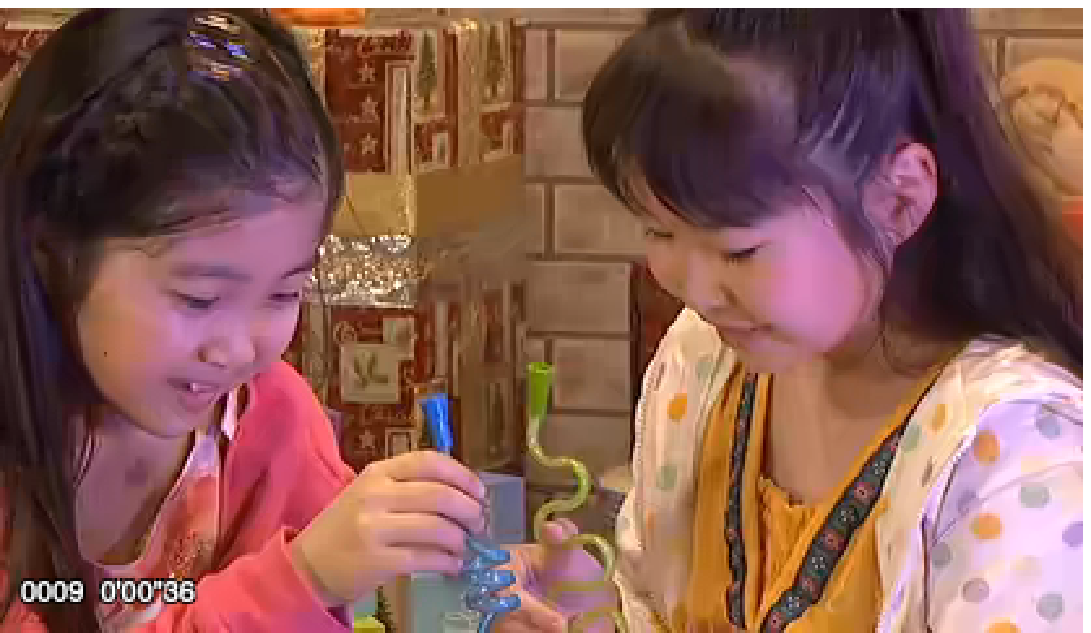}}
\subfigure[$\mathbf{\hat{C}}_9$]{\includegraphics[scale=.54]{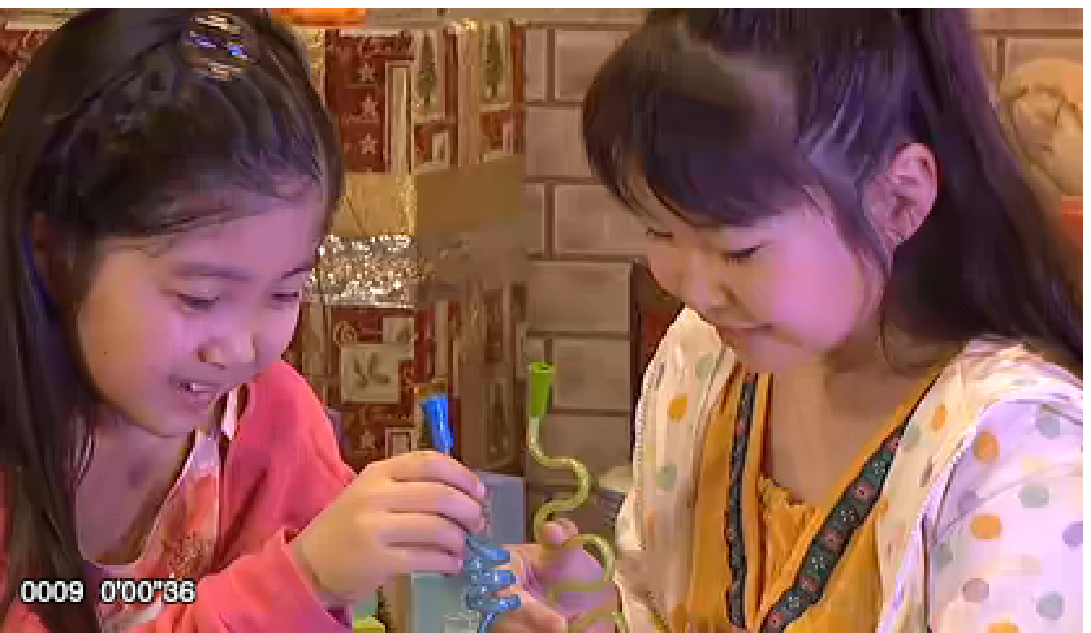}}\\
\caption{
Compressed frame of ``BlowingBubbles'' using AI coding configuration with $\mbox{QP} = 32$. Core transforms are the default HEVC DCT, and the low-complexity DCT approximations.}
\label{fig:exemplehevc}
\end{figure}

For qualitative assessment, Figure~\ref{fig:exemplehevc} illustrates the tenth frame of the ``BlowingBubbles'' video sequence using the original HEVC configurations and the modified versions of the reference software after embedding the discussed \mbox{8-,} 16-, and 32-point DCT approximations.
Such results consider the main AI configuration mode and the quantization parameter (QP) set to $32$.

The video frame in Figure~\ref{fig:exemplehevc} show that
any existing visual degradation are essentially imperceptible;
and therefore
the switching from the original HEVC transforms to the proposed multiparametric low-complexity DCT approximations
does not result in significant losses.
Table~\ref{T:exemplehevc}
summarizes the quantitative measurements related to
the video frame shown in Figure~\ref{fig:exemplehevc}.
One may note that for the given video frame,
our modified HEVC reference software may lead to better results than the original one
in terms of MSE and PSNR depending on the color channel.

\begin{table}
\setlength{\tabcolsep}{4pt}
\centering
\caption{MSE and PSNR measurements for compressed frame of ``BlowingBubbles'' using AI coding configuration with $\mbox{QP} = 32$.}
\label{T:exemplehevc}
\begin{tabular}{ccccc}
\hline
& HEVC DCT & $\mathbf{\hat{C}}_1$ & $\mathbf{\hat{C}}_5$ & $\mathbf{\hat{C}}_9$ \\
\hline
$\mbox{MSE-Y}$  & $30.957$ & $31.229$ & $31.065$ & $30.763$ \\
$\mbox{MSE-U}$  & $13.793$ & $13.769$ & $13.689$ & $14.092$ \\
$\mbox{MSE-V}$  &  $9.934$ & $10.187$ & $10.006$ & $9.731$  \\
$\mbox{PSNR-Y}$ (dB) & $33.223$ & $33.175$ & $33.201$ & $33.250$ \\
$\mbox{PSNR-U}$ (dB) & $36.734$ & $36.742$ & $36.767$ & $36.641$ \\
$\mbox{PSNR-V}$ (dB) & $38.159$ & $38.050$ & $38.128$ & $38.249$ \\
\hline
\end{tabular}
\end{table}

	\section{Conclusion}

\label{section-conclusion}

	In this work, we proposed
a novel class of low-complexity transforms based on
the mathematical frameworks introduced by~\cite{bas2008,bas2009,bas2010,bas2011,bas2013}.
	We have presented a parametrized fast algorithm
linked to the introduced 8-point DCT approximations.
Based on a multicriteria optimization scheme,
we could jointly
optimize
the coding efficiency metrics and the distance between the candidate transforms to the exact DCT
in order to identify the
best-performing 8-point approximations.
As a result, we obtained fifteen optimal DCT approximations,
which were submitted to the JAM scaling method
for obtaining novel 16- and 32-point transforms.

The obtained 8-point and scaled
transforms were assessed
in terms of computational complexity,
coding efficiency,
and
similarity to the DCT.
Furthermore, we submitted the introduced
8-, 16- and 32-point transforms to both image and videos coding experiments
along
with
extensive
comparisons with
competing
low-complexity DCT approximations.
Results show that the proposed DCT approximations
can outperform several peering methods
in terms of image and video coding
quality metrics.
We emphasize that
the very low complexity of the proposed transforms can be fundamental for
efficient hardware implementation in scenarios of
limited resources (energy autonomy or processing power) and real-time encoding,
as illustrated in the context of
wireless visual sensor networks.

Further research in this area can be pursued
in terms of deriving low-complexity methods
for
non-trigonometric transforms,
such as discrete wavelet transforms.
However,
discrete wavelets constitute a
fundamentally different approach when compared to
usual trigonometric transforms,
such as the DCT.
Therefore, new tools and comparison
procedures are expected to be derived.

\section*{Acknowledgments}

We gratefully
acknowledge partial financial support from
Funda\c{c}\~ao de Amparo a Ci\^encia e Tecnologia do Estado de Pernambuco (FACEPE),
Conselho Nacional de Desenvolvimento Cient\'{\i}fico and Tecnol\'ogico (CNPq),
and Coordena\c{c}\~ao de Aperfei\c{c}oamento de Pessoal de N\'ivel Superior (CAPES),
Brazil.

{\small
\singlespacing
\bibliographystyle{IEEEtran}
\bibliography{dct}

\begin{thebibliography}{10}
\providecommand{\url}[1]{#1}
\csname url@samestyle\endcsname
\providecommand{\newblock}{\relax}
\providecommand{\bibinfo}[2]{#2}
\providecommand{\BIBentrySTDinterwordspacing}{\spaceskip=0pt\relax}
\providecommand{\BIBentryALTinterwordstretchfactor}{4}
\providecommand{\BIBentryALTinterwordspacing}{\spaceskip=\fontdimen2\font plus
\BIBentryALTinterwordstretchfactor\fontdimen3\font minus
  \fontdimen4\font\relax}
\providecommand{\BIBforeignlanguage}[2]{{%
\expandafter\ifx\csname l@#1\endcsname\relax
\typeout{** WARNING: IEEEtran.bst: No hyphenation pattern has been}%
\typeout{** loaded for the language `#1'. Using the pattern for}%
\typeout{** the default language instead.}%
\else
\language=\csname l@#1\endcsname
\fi
#2}}
\providecommand{\BIBdecl}{\relax}
\BIBdecl

\bibitem{Ahmed1974}
N.~Ahmed, T.~Natarajan, and K.~Rao, ``Discrete cosine transfom,'' \emph{IEEE
  Transactions on Computers}, vol.~23, no.~1, pp. 90--93, 1974.

\bibitem{Ahmed1975}
N.~Ahmed and K.~R. Rao, \emph{Orthogonal Transforms for Digital Signal
  Processing}.\hskip 1em plus 0.5em minus 0.4em\relax Berlin: Springer-Verlag,
  1975.

\bibitem{Oppenheim2009}
A.~V. Oppenheim and R.~W. Schafer, \emph{Discrete-Time Signal Processing},
  3rd~ed.\hskip 1em plus 0.5em minus 0.4em\relax Upper Saddle River, NJ, USA:
  Prentice Hall Press, 2009.

\bibitem{Efros2004}
M.~Effros, H.~Feng, and K.~Zeger, ``Suboptimality of the {K}arhunen-{L}o{\`e}ve
  transform for transform coding,'' \emph{IEEE Transactions on Information
  Theory}, vol.~50, no.~8, pp. 1605--1619, Aug. 2004.

\bibitem{Cla1981}
R.~J. Clark, ``Relation between {K}arhunen-{L}o\`eve and cosine transform,'' in
  \emph{IEE Proceedings-F, Communications, Radar and Signal Processing}, vol.
  128, no.~6, 1981, pp. 359--360.

\bibitem{Gonzalez2008}
R.~C. Gonzalez and R.~E. Woods, \emph{Digital image processing}.\hskip 1em plus
  0.5em minus 0.4em\relax Upper Saddle River, N.J.: Prentice Hall, 2008.

\bibitem{Home}
\BIBentryALTinterwordspacing
JPEG, ``Joint photographic experts group,'' 2012. [Online]. Available:
  \url{http://www.jpeg.org}
\BIBentrySTDinterwordspacing

\bibitem{Wiegand2003}
T.~Wiegand, G.~J. Sullivan, G.~Bj{\o}ntegaard, and A.~Luthra, ``Overview of the
  {H}.264/{AVC} video coding standard,'' \emph{IEEE Transactions on Circuits
  and Systems for Video Technology}, vol.~13, no.~7, pp. 560--576, Jul. 2003.

\bibitem{hevc1}
M.~T. Pourazad, C.~Doutre, M.~Azimi, and P.~Nasiopoulos, ``{HEVC}: The new gold
  standard for video compression: How does {HEVC} compare with {H.264/AVC}?''
  \emph{IEEE Consumer Electronics Magazine}, vol.~1, no.~3, pp. 36--46, Jul.
  2012.

\bibitem{Chen1977}
W.~H. Chen, C.~Smith, and S.~Fralick, ``A fast computational algorithm for the
  discrete cosine transform,'' \emph{IEEE Transactions on Communications},
  vol.~25, no.~9, pp. 1004--1009, 1977.

\bibitem{Lee1984}
B.~G. Lee, ``A new algorithm for computing the discrete cosine transform,''
  \emph{IEEE Transactions on Acoustics, Speech and Signal Processing}, vol.
  ASSP-32, pp. 1243--1245, Dec. 1984.

\bibitem{Loef1989}
C.~Loeffler, A.~Ligtenberg, and G.~Moschytz, ``Practical fast 1{D} {DCT}
  algorithms with 11 multiplications,'' in \emph{Proceedings of the
  International Conference on Acoustics, Speech, and Signal Processing}, 1989,
  pp. 988--991.

\bibitem{Feig1992}
F.~Feig and S.~Winograd, ``Fast algorithms for the discrete cosine transform,''
  \emph{IEEE Transactions on Signal Processing}, vol.~40, no.~9, pp.
  2174--2193, 1992.

\bibitem{Brita2007}
V.~Britanak, P.~Yip, and K.~R. Rao, \emph{Discrete Cosine and Sine
  Transforms}.\hskip 1em plus 0.5em minus 0.4em\relax Academic Press, 2007.

\bibitem{Wang2011}
Z.~Wang and A.~C. Bovik, ``Reduced- and no-reference image quality
  assessment,'' \emph{IEEE Signal Processing Magazine}, vol.~28, no.~6, pp.
  29--40, 2011.

\bibitem{Saponara2012}
S.~Saponara, ``Real-time and low-power processing of 3{D} direct/inverse
  discrete cosine transform for low-complexity video codec,'' \emph{Journal of
  Real-Time Image Processing}, vol.~7, no.~1, pp. 43--53, Mar. 2012.

\bibitem{bas2013}
S.~Bouguezel, M.~O. Ahmad, and M.~N.~S. Swamy, ``Binary discrete cosine and
  {H}artley transforms,'' \emph{IEEE Transactions on Circuits and Systems I:
  Regular Papers}, vol.~60, no.~4, pp. 989--1002, 2013.

\bibitem{Harize2013}
N.~{Kouadria}, N.~{Doghmane}, D.~{Messadeg}, and S.~{Harize}, ``Low complexity
  {DCT} for image compression in wireless visual sensor networks,''
  \emph{Electronics Letters}, vol.~49, no.~24, pp. 1531--1532, Nov. 2013.

\bibitem{Kouadria2017}
N.~Kouadria, K.~Mechouek, D.~Messadeg, and N.~Doghmane, ``Pruned discrete
  {T}chebichef transform for image coding in wireless multimedia sensor
  networks,'' \emph{AEU - International Journal of Electronics and
  Communications}, vol.~74, pp. 123--127, 2017.

\bibitem{Haweel2001}
T.~I. Haweel, ``A new square wave transform based on the {DCT},'' \emph{Signal
  Processing}, vol.~82, pp. 2309--2319, 2001.

\bibitem{Bayer2012}
F.~M. Bayer and R.~J. Cintra, ``{DCT}-like transform for image compression
  requires 14 additions only,'' \emph{Electronics Letters}, vol.~48, no.~15,
  pp. 919--921, 2012.

\bibitem{cb2011}
R.~J. Cintra and F.~M. Bayer, ``A {DCT} approximation for image compression,''
  \emph{IEEE Signal Processing Letters}, vol.~18, no.~10, pp. 579--582, 2011.

\bibitem{Tablada2015}
C.~J. Tablada, F.~M. Bayer, and R.~J. Cintra, ``A class of {DCT} approximations
  based on the {F}eig-{W}inograd algorithm,'' \emph{Signal Processing (Print)},
  vol.~11, pp. 1--20, 2015.

\bibitem{bas2011}
S.~Bouguezel, M.~O. Ahmad, and M.~N.~S. Swamy, ``A low-complexity parametric
  transform for image compression,'' in \emph{Proceedings of the 2011 IEEE
  International Symposium on Circuits and Systems}, 2011, pp. 2145--2148.

\bibitem{multibeam2012}
U.~S. Potluri, A.~Madanayake, R.~J. Cintra, F.~M. Bayer, and N.~Rajapaksha,
  ``Multiplier-free {DCT} approximations for {RF} multi-beam digital
  aperture-array space imaging and directional sensing,'' \emph{Measurement
  Science and Technology}, vol.~23, no.~11, p. 114003, 2012.

\bibitem{Potluri2013}
U.~S. Potluri, A.~Madanayake, R.~J. Cintra, F.~M. Bayer, S.~Kulasekera, and
  A.~Edirisuriya, ``Improved 8-point approximate {DCT} for image and video
  compression requiring only 14 additions,'' \emph{IEEE Transactions on
  Circuits and Systems I}, vol.~61, no.~6, pp. 1727 -- 1740, 2013.

\bibitem{Bayer2010}
F.~M. Bayer and R.~J. Cintra, ``Image compression via a fast {DCT}
  approximation,'' \emph{IEEE Latin America Transactions}, vol.~8, no.~6, pp.
  708--713, Dec. 2010.

\bibitem{bayer201216pt}
F.~M. Bayer, R.~J. Cintra, A.~Edirisuriya, and A.~Madanayake, ``A digital
  hardware fast algorithm and {FPGA}-based prototype for a novel 16-point
  approximate {DCT} for image compression applications,'' \emph{Measurement
  Science and Technology}, vol.~23, no.~8, pp. 114\,010--114\,019, 2012.

\bibitem{Bayer2013}
F.~M. Bayer, R.~J. Cintra, A.~Madanayake, and U.~S. Potluri, ``Multiplierless
  approximate 4-point {DCT} {VLSI} architectures for transform block coding,''
  \emph{Electronics Letters}, vol.~49, no.~24, pp. 1532--1534, Nov. 2013.

\bibitem{cintra2014low}
R.~J. Cintra, F.~M. Bayer, and C.~J. Tablada, ``Low-complexity 8-point {DCT}
  approximations based on integer functions,'' \emph{Signal Processing},
  vol.~99, pp. 201--214, 2014.

\bibitem{bas2008}
S.~Bouguezel, M.~O. Ahmad, and M.~N.~S. Swamy, ``Low-complexity 8$\times$8
  transform for image compression,'' \emph{Electronics Letters}, vol.~44,
  no.~21, pp. 1249--1250, Sep. 2008.

\bibitem{bas2008b}
------, ``A multiplication-free transform for image compression,'' in \emph{2nd
  International Conference on Signals, Circuits and Systems}, Nov. 2008, pp.
  1--4.

\bibitem{bas2009}
------, ``A fast 8$\times$8 transform for image compression,'' in \emph{2009
  International Conference on Microelectronics (ICM)}, Dec. 2009, pp. 74--77.

\bibitem{bas2010}
------, ``A novel transform for image compression,'' in \emph{53rd IEEE
  International Midwest Symposium on Circuits and Systems (MWSCAS)}, Aug. 2010,
  pp. 509--512.

\bibitem{Oliveira2019}
R.~S. Oliveira, R.~J. Cintra, F.~M. Bayer, T.~L.~T. da~Silveira, A.~Madanayake,
  and A.~Leite, ``Low-complexity 8-point {DCT} approximation based on angle
  similarity for image and video coding,'' \emph{Multidimensional Systems and
  Signal Processing}, vol.~30, no.~3, pp. 1363--1394, Jul. 2019.

\bibitem{Coutinho2017}
V.~A. Coutinho, R.~J. Cintra, and F.~M. Bayer, ``Low-complexity
  multidimensional {DCT} approximations for high-order tensor data
  decorrelation,'' \emph{IEEE Transactions on Image Processing}, vol.~26,
  no.~5, pp. 2296--2310, 2017.

\bibitem{Ezhilarasi2018}
R.~Ezhilarasi, K.~Venkatalakshmi, and B.~Khanth, ``Enhanced approximate
  discrete cosine transforms for image compression and multimedia
  applications,'' \emph{Multimedia Tools and Applications}, vol. (Online
  First), pp. 1--14, 2018.

\bibitem{Almurib2018}
H.~A.~F. {Almurib}, T.~N. {Kumar}, and F.~{Lombardi}, ``Approximate {DCT} image
  compression using inexact computing,'' \emph{IEEE Transactions on Computers},
  vol.~67, no.~2, pp. 149--159, 2018.

\bibitem{Zhang2019}
J.~Zhang, W.~Shi, L.~Zhou, R.~Gong, L.~Wang, and H.~Zhou, ``A low-power and
  high-{PSNR} unified {DCT}/{IDCT} architecture based on {EARC} and enhanced
  scale factor approximation,'' \emph{IEEE Access}, vol.~7, pp.
  165\,684--165\,691, 2019.

\bibitem{Huang2019}
J.~{Huang}, T.~{Nandha Kumar}, H.~A.~F. {Almurib}, and F.~{Lombardi}, ``A
  deterministic low-complexity approximate (multiplier-less) technique for
  {DCT} computation,'' \emph{IEEE Transactions on Circuits and Systems I:
  Regular Papers}, vol.~66, no.~8, pp. 3001--3014, 2019.

\bibitem{Zidani2019}
N.~{Zidani}, N.~{Kouadria}, N.~{Doghmane}, and S.~{Harize}, ``Low complexity
  pruned {DCT} approximation for image compression in wireless multimedia
  sensor networks,'' in \emph{2019 5th International Conference on Frontiers of
  Signal Processing (ICFSP)}, Sep. 2019, pp. 26--30.

\bibitem{Brahimi2020}
N.~Brahimi, T.~Bouden, T.~Brahimi, and L.~Boubchir, ``A novel and efficient
  8-point {DCT} approximation for image compression,'' \emph{Multimedia Tools
  and Applications}, vol. (Online First), pp. 1--17, 2020.

\bibitem{Ohm2012}
J.-R. Ohm, G.~J. Sullivan, H.~Schwarz, T.~K. Tan, and T.~Wiegand, ``Comparison
  of the coding efficiency of video coding standards - including high
  efficiency video coding ({HEVC}),'' \emph{IEEE Transactions on Circuits and
  Systems for Video Technology}, vol.~22, no.~12, pp. 1669--1684, Dec. 2012.

\bibitem{Sullivan2012}
G.~J. Sullivan, J.~R. Ohm, W.~J. Han, and T.~Wiegand, ``Overview of the high
  efficiency video coding ({HEVC}) standard,'' \emph{IEEE Transactions on
  Circuits and Systems for Video Technology}, vol.~22, no.~12, pp. 1649--1668,
  2012.

\bibitem{Jrid2015}
M.~Jridi, A.~Alfalou, and P.~K. Meher, ``A generalized algorithm and
  reconfigurable architecture for efficient and scalable orthogonal
  approximation of {DCT},'' \emph{IEEE Transactions on Circuits and Systems I:
  Regular Papers}, vol.~62, no.~2, pp. 449--457, 2015.

\bibitem{Wallace1992}
G.~K. Wallace, ``The {JPEG} still picture compression standard,'' \emph{IEEE
  Transactions on Consumer Electronics}, vol.~30, no.~1, pp. xviii--xxxiv,
  1992.

\bibitem{Leng2004}
K.~Lengwehasatit and A.~Ortega, ``Scalable variable complexity approximate
  forward {DCT},'' \emph{IEEE Transactions on Circuits and Systems for Video
  Technology}, vol.~14, no.~11, pp. 1236--1248, Nov. 2004.

\bibitem{Cintra2011}
R.~J. Cintra, ``An integer approximation method for discrete sinusoidal
  transforms,'' \emph{Circuits, Systems, and Signal Processing}, vol.~30, pp.
  1481--1501, 2011.

\bibitem{Higham1987}
N.~J. Higham, ``Computing real square roots of a real matrix,'' \emph{Linear
  Algebra and its Applications}, vol. 88-89, pp. 405--430, 1987.

\bibitem{jayant1984digital}
N.~S. Jayant and P.~Noll, \emph{Digital Coding of Waveforms, Principles and
  Applications to Speech and Video}.\hskip 1em plus 0.5em minus 0.4em\relax
  Englewood Cliffs NJ, USA: Prentice-Hall, 1984, p. 688.

\bibitem{R2016}
\BIBentryALTinterwordspacing
R.~{C}ore {T}eam, \emph{{R}: {A} {L}anguage and {E}nvironment for {S}tatistical
  {C}omputing}, {R} {F}oundation for {S}tatistical {C}omputing, {V}ienna,
  {A}ustria, 2016. [Online]. Available: \url{https://www.R-project.org/}
\BIBentrySTDinterwordspacing

\bibitem{Oliveira2013b}
R.~S. Oliveira, R.~J. Cintra, F.~M. Bayer, and C.~J. Tablada, ``Uma
  aproxima\c{c}\~ao ortogonal para a {DCT},'' in \emph{XXXI Simp\'osio
  Brasileiro de Telecomunica\c{c}\~oes}, 2013.

\bibitem{Brahimi2011}
N.~Brahimi and S.~Bouguezel, ``An efficient fast integer {DCT} transform for
  images compression with 16 additions only,'' in \emph{Internatinal Workshop
  on Systems, Signal Processing and their Applications}, 2011, 71-74.

\bibitem{Miettinen1999}
K.~Miettinen, \emph{Nonlinear Multiobjective Optimization}, ser. International
  Series in Operations Research and Management Science.\hskip 1em plus 0.5em
  minus 0.4em\relax Kluwer Academic Publishers, Dordrecht, 1999, vol.~12.

\bibitem{Silveira201660}
T.~L.~T. da~Silveira, F.~M. Bayer, R.~J. Cintra, S.~Kulasekera, A.~Madanayake,
  and A.~J. Kozakevicius, ``An orthogonal 16-point approximate {DCT} for image
  and video compression,'' \emph{Multidimensional Systems and Signal
  Processing}, vol.~27, no.~1, pp. 87--104, 2016.

\bibitem{Silveira201644}
T.~L.~T. da~Silveira, R.~S. Oliveira, F.~M. Bayer, R.~J. Cintra, and
  A.~Madanayake, ``Multiplierless 16-point {DCT} approximation for
  low-complexity image and video coding,'' \emph{Signal, Image and Video
  Processing}, vol.~11, no.~2, pp. 227--233, 2017.

\bibitem{Base}
\BIBentryALTinterwordspacing
(2011) {USC}-{SIPI}: The {USC}-{SIPI} {I}mage {D}atabase. University of
  Southern California, Signal and Image Processing Institute.
  http://sipi.usc.edu/database/. {U}niversity of {S}outhern {C}alifornia,
  {S}ignal and {I}mage {P}rocessing {I}nstitute. [Online]. Available:
  \url{http://sipi.usc.edu/database/}
\BIBentrySTDinterwordspacing

\bibitem{Thu2008}
Q.~H. Thu and M.~Ghanbari, ``Scope of validity of {PSNR} in image/video quality
  assessment.'' \emph{Electronics Letters}, vol.~44, no.~13, pp. 800--801,
  2008.

\bibitem{Wang2004}
Z.~Wang, A.~C. Bovik, H.~R. Sheikh, and E.~P. Simoncelli, ``Image quality
  assessment: from error visibility to structural similarity,'' \emph{IEEE
  Transactions on Image Processing}, vol.~13, no.~4, pp. 600--612, 2004.

\bibitem{Hig2008}
N.~J. Higham, \emph{Functions of matrices: theory and computation}.\hskip 1em
  plus 0.5em minus 0.4em\relax SIAM, 2008, vol. 104.

\bibitem{refsoft}
\BIBentryALTinterwordspacing
{Joint Collaborative Team on Video Coding (JCT-VC)}, ``{HEVC} reference
  software documentation,'' 2013, {F}raunhofer {H}einrich {H}ertz Institute.
  [Online]. Available: \url{https://hevc.hhi.fraunhofer.de/}
\BIBentrySTDinterwordspacing

\bibitem{CTConditions2013}
F.~Bossen, ``Common test conditions and software reference configurations,''
  San Jose, CA, USA, Feb. 2013, document JCT-VC L1100.

\bibitem{Bjontegaard2001}
G.~Bj{\o}ntegaard, ``Calculation of average {PSNR} differences between
  {RD}-curves,'' in \emph{13th VCEG Meeting}, Austin, TX, USA, Apr. 2001,
  document VCEG-M33.

\bibitem{Hanhart2014}
P.~Hanhart and T.~Ebrahimi, ``Calculation of average coding efficiency based on
  subjective quality scores,'' \emph{Journal of Visual Communication and Image
  Representation}, vol.~25, no.~3, pp. 555 -- 564, 2014.

\end{thebibliography}
}

\end{document}